\documentclass[sigconf]{acmart}

\usepackage{amsfonts}

\usepackage{amssymb,amsmath}
\usepackage{color}
\usepackage{appendix}
\usepackage[noend]{algpseudocode}
\usepackage{algorithmicx,algorithm}
\usepackage{amsmath}  

\newcommand{\ignore}[1]{}
\usepackage{xspace}
\newcommand{\paratitle}[1]{\vspace{1.5ex}\noindent\textbf{#1}}

\usepackage{multirow}
\usepackage{multicol}
\usepackage{caption}
\usepackage{subcaption}
\usepackage{latexsym}
\usepackage{mflogo}
\usepackage{enumitem}
\usepackage{marvosym}

\AtBeginDocument{%
  \providecommand\BibTeX{{%
    \normalfont B\kern-0.5em{\scshape i\kern-0.25em b}\kern-0.8em\TeX}}}

\setcopyright{acmcopyright}

\copyrightyear{2022}
\acmYear{2022}
\setcopyright{acmcopyright}
\acmConference[SIGIR '22]{Proceedings of the 45th International ACM SIGIR Conference on Research and Development in Information Retrieval}{July 11--15, 2022}{Madrid, Spain.}
\acmBooktitle{Proceedings of the 45th International ACM SIGIR Conference on Research and Development in Information Retrieval (SIGIR '22), July 11--15, 2022, Madrid, Spain}
\acmPrice{15.00}
\acmISBN{978-1-4503-8732-3/22/07}
\acmDOI{10.1145/3477495.3531931}

\usepackage{xcolor}

\begin{document}

\title{Ada-Ranker: A Data Distribution Adaptive Ranking Paradigm for Sequential Recommendation}


\author{Xinyan Fan}
\email{xinyan.fan@ruc.edu.cn}
\affiliation{
	\institution{Gaoling School of Artificial Intelligence, Renmin University of China}
	\institution{Beijing Key Laboratory of Big Data Management and Analysis Methods}
	\city{Beijing}
	\country{China}
}

\author{Jianxun Lian}
\email{jianxun.lian@microsoft.com}
\affiliation{
	\institution{Microsoft Research Asia}
	\city{Beijing}
	\country{China}
}

\author{Wayne Xin Zhao\textsuperscript{\Letter}} \thanks{\textsuperscript{\Letter} Corresponding author.} 
\email{batmanfly@gmail.com}
\affiliation{
	\institution{Gaoling School of Artificial Intelligence, Renmin University of China}
	\institution{Beijing Key Laboratory of Big Data Management and Analysis Methods}
	\institution{Beijing Academy of Artificial Intelligence}
	\city{Beijing}
	\country{China}
}

\author{Zheng Liu}
\email{zhengliu@microsoft.com}
\affiliation{
	\institution{Microsoft Research Asia}
	\city{Beijing}
	\country{China}
}

\author{Chaozhuo Li}
\email{cli@microsoft.com}
\affiliation{
	\institution{Microsoft Research Asia}
	\city{Beijing}
	\country{China}
}

\author{Xing Xie}
\email{xingx@microsoft.com}
\affiliation{
	\institution{Microsoft Research Asia}
	\city{Beijing}
	\country{China}
}
\renewcommand{\authors}{Xinyan Fan, Jianxun Lian, Wayne Xin Zhao, Zheng Liu, Chaozhuo Li and Xing Xie}

\renewcommand{\shortauthors}{Xinyan Fan, Jianxun Lian, Wayne Xin Zhao, Zheng Liu, Chaozhuo Li and Xing Xie}



\begin{abstract}
A large-scale recommender system usually consists of \emph{recall} and \emph{ranking} modules. The goal of ranking modules (aka \emph{rankers}) is to elaborately discriminate users' preference on item candidates proposed by recall modules. With the success of deep learning techniques in various domains, we have witnessed the mainstream rankers evolve from traditional models to deep neural models. However, the way that we design and use rankers remains unchanged: offline training the model, freezing the parameters, and deploying it for online serving. Actually, the candidate items are determined by specific user requests, in which underlying distributions (e.g., the proportion of items for different categories, the proportion of popular or new items) are highly different from one another in a production environment. The classical parameter-frozen inference manner cannot adapt to dynamic serving circumstances, making rankers' performance compromised.

In this paper, we propose a new training and inference paradigm, termed as \emph{Ada-Ranker}, to address the challenges of dynamic online serving. Instead of using parameter-frozen models for universal serving, Ada-Ranker can adaptively modulate parameters of a ranker according to the data distribution of the current group of item candidates. We first extract distribution patterns from the item candidates. Then, we modulate the ranker by the patterns to make the ranker adapt to the current data distribution. Finally, we use the revised ranker to score the candidate list. In this way, we empower the ranker with the capacity of adapting from a global model to a local model which better handles the current task. As a first study, we examine our Ada-Ranker paradigm in the sequential recommendation scenario.  Experiments on three datasets demonstrate that Ada-Ranker can effectively enhance various base sequential models and also outperform a comprehensive set of competitive baselines.

\end{abstract}

\ccsdesc[500]{Information systems~Learning to rank}
\ccsdesc[500]{Information systems~Recommender systems}
\ccsdesc[300]{Computing methodologies~Neural networks}
\keywords{Model Adaptation, Sequential Recommendation, Dynamic Ranking}

\maketitle

\section{introduction} 
\label{sec:intro}
Recommender systems play an important role in information filtering for online services such as e-commerce, news, movie, and gaming. To support efficient recommendations from a massive set of items, an industrial recommender system usually follows the \textsl{recall-then-rank} two-stage paradigm. Given a user request, the recall component proposes a small set of relevant candidates with lightweight methods, then the ranking component further elaborately scores the candidates with more advanced models and returns top-$k$ results. This paper discusses models in the ranking component (which we call \textbf{\textsl{rankers}} hereafter). 

Since online user behaviors are highly dynamic (driven by evolving user preference and short-term interest), enhancing rankers with sequential user modeling becomes a hot research topic and shows significant business value in industry~\cite{Zhou_Mou_Fan_Pi_Bian_Zhou_Zhu_Gai_2019,lian2021multi}.  In recent literature, a number of sequential recommendation models have been proposed based on neural network architectures, 
including Gated Recurrent Unit (GRU)~\cite{DBLP:conf/emnlp/ChoMGBBSB14}, Convolutional Neural Network~(CNN)~\cite{NIPS2012_c399862d} and Transformers~\cite{10.5555/3295222.3295349}. These approaches 
 can effectively handle sequential interaction data and train the recommendation models in an end-to-end manner. 
\ignore{In recent years, the research interests in rankers have switched from feature-based models, such as FM~\cite{DBLP:conf/icdm/Rendle10} and GBDT~\cite{10.1145/2939672.2939785}, to deep learning-based models, such as  DeepFM~\cite{10.5555/3172077.3172127} and DIN~\cite{10.1145/3219819.3219823}. Deep neural networks have advantages in automatically learning useful patterns from complex raw data, such as high-order feature interactions~\cite{10.1145/3219819.3220023}, dynamic attentions~\cite{10.1145/3219819.3219823} and efficient model transferring~\cite{10.1145/3397271.3401156}, without the requirement of exhausting feature engineering effort. 
Take the sequential recommendation for example, users' behaviors are ordered by timestamps and its task is to recommend items that user will interact in the near future. It is hard to manually extract a comprehensive set of features to represent the dependency of user's actions and user's interest evolution from the behavior history. In contrast, there are a few deep learning modules, such as Gated Recurrent Unit (GRU)~\cite{DBLP:conf/emnlp/ChoMGBBSB14}, Convolutional Neural Network~(CNN)~\cite{NIPS2012_c399862d} and Transformers~\cite{10.5555/3295222.3295349}, which can handle sequential data and train a model in an end-to-end manner.
}
Typically, these neural rankers still follow the ``\textsl{regular offline training $\Rightarrow$ static online inference}" paradigm, in which 
once a model is trained, the parameters are frozen and deployed into online environment for instant service. 

However, the candidate lists to be ranked are determined by specific user requests, and the underlying distributions can be rather different in diverse request scenarios. 
For example, in news recommendations, for users who are big fans of NBA, their recall candidates will contain more news articles related to basketball than other users; on an e-commerce platform, if a user clicks on the \textsl{Women's Fashion} category, then the retrieved candidates will be highly related to this category. Besides,  other factors (e.g., temporal effects~\cite{10.1145/3097983.3098096} and counterfactual settings~\cite{10.1145/3383313.3411552})
might also increase the discrepancy among the data distributions of candidate items for different user requests. 
Therefore, the parameter-frozen inference paradigm for rankers can only produce  sub-optimal performances since data distribution discrepancy will cause significant performance decrease in the rankers. Indeed, such an issue generally exists in supervised learning methods~\cite{pmlr-v97-recht19a,pmlr-v119-sun20b}. It is desirable to develop a more  capable ranking paradigm  that can flexibly adapt to different circumstances.

As existing solutions, several methods curate a few online circumstances (e.g., for different weekdays~\cite{10.1145/3097983.3098096} or different domains~\cite{sheng2021model}), and then let the model dynamically select a suitable one for response. However, these approaches can only cover a limited set of recommendation scenarios by assuming they are pre-given, and we need to maintain multiple models in the service system, which is not parameter-efficient.
Besides, it takes a significant cost to switch among different heavy model variants, which  also increases the risk of system failures. Therefore, it is not realistic to assume that the circumstances are discrete and enumerable. It is still challenging to design a capable ranker that dynamically adapts to a specific circumstance in a  flexible and parameter-efficient manner.
  
\ignore{these approaches have many drawbacks: first, it can only cover a very limited set of circumstances, and it assumes that all of them are known beforehand; second, maintaining multiple models in a serving system is not parameter-efficient, what is worse, frequent switching between models will increase the risk of system failures; third, it assumes that circumstances are discrete and enumerable, which does not fit the real-world cases where a ranking request can be a soft mixture of multiple conditions.  How to make a ranker dynamically adapt to a specific circumstance in a comprehensive, flexible, and parameter-efficient manner remains an open question.
}

In this paper, we propose \textbf{Ada-Ranker} -- a new \underline{Ada}ptive paradigm for making context-aware inference in \underline{Rankers}.  We address the data distribution shift issue by a \emph{model adaptation} approach. Specifically, we
regard handling each set of item candidates from recall modules to a specific user request as an individual \textsl{task}. During the inference stage for a task, instead of fixing the parameters of a ranker, Ada-Ranker adapts the ranker to the current task in three steps, namely 
\emph{distribution learning}, 
\emph{input modulation} and \emph{parameter modulation}. For distribution learning, we learn the underlying data patterns of the current task by Neural Processes~\cite{garnelo2018neural,garnelo2018conditional}, which will help the ranker focus on extracting useful users behavior patterns to better discriminate candidate items in current task. For input modulation, by taking the extracted data distribution patterns as adaptation conditions, we learn a feature-wise linear modulation neural network to adjust the input representations, so that input representations are re-located to latent positions where rankers can discriminate the current task better. For parameter modulation, we adopt a model patching approach by generating parameter patches based on a parameter pool of base parameter vectors or matrices. 
Our adaptation process consists of the above input modulation and parameter modulation procedures, which jointly adapt model's parameters according to specific tasks in a model-agnostic way. 
To verify the effectiveness of Ada-Ranker, we design comprehensive settings for test environments and apply Ada-Ranker to various types of base sequential recommendation models on three real-world datasets. Experiment results demonstrate that Ada-Ranker can effectively adapt to dynamic and diverse test tasks, and significantly boost the base models.

\ignore{Firstly, it learns the data distribution patterns of the target task and utilizes  it to adjust the ranker, so that the ranker can focus on discriminating the user-item patterns in current input. Secondly, it utilizes  the revised ranker to score the candidates, which is the same as the scoring process of traditional rankers. Obviously, the major two challenges for Ada-Ranker lie in its first step: how to represent the data distribution of a given group of candidates and how to effectively perform model adaptation. All these operations should be parameter-efficient and light-weighted, so that the adaptation process will not bring too many burdens to the original model.
}
\ignore{To this end, we design several mechanisms in Ada-Ranker. Firstly, we leverage neural process to encode all candidates, and obtain the data distribution pattern of current task. 
Secondly, we propose a novel adaptation network to conduct model adaptation. The adaptation process includes the input modulation and the parameter modulation, which are model-agnostic and work jointly to revise model's parameters. 
Thirdly, we pre-train the base ranker with the traditional supervised learning paradigm, and only update adaptation parameters of Ada-Ranker. This efficient training strategy makes our model plug-and-play.}

Our contributions are summarized as follows. 
(1) We highlight the data distribution discrepancy issue and the importance of parameter adaptation during the inference process, which is commonly encountered in industrial recommender systems but overlooked in literature. 
(2) We propose Ada-Ranker, a new inference paradigm for rankers, which can adaptively adjust the model parameters according to the data patterns in a given set of item candidates. Ada-Ranker satisfies the three properties of being lightweight, model-agnostic and flexible for plug-and-play usage.
(3) We conduct extensive experiments on three real-world datasets in the scenario of sequential recommendations. Results demonstrate that Ada-Ranker can significantly boost the performance for various base models, as well as outperform a set of competitive baselines.

\ignore{Our main contributions are summarized as follows:
\begin{enumerate}[leftmargin=*]
	\item We highlight the data distribution inconsistency issue and the importance of parameter adaptation during the inference process, which is commonly encountered in industrial recommender systems but overlooked in literature.
	\item We propose Ada-Ranker, a new inference paradigm for rankers, which can adaptively adjust the model parameters according to the patterns in a given group of item candidates. Ada-Ranker satisfies the three properties of being light-weighted, model-agnostic and flexible for plug-and-play usage.
	\item We conduct extensive experiments on three real-world datasets in the scenario of sequential recommendations. Results demonstrate that Ada-Ranker can significantly boost the performance for various base models, as well as outperform a comprehensive set of competitive baselines.
\end{enumerate}
}

\section{Problem Formulation}

Given a request from a user $u$, the recall module uses multiple approaches (such as popularity-based, item-to-item and approximate nearest neighbor search) to retrieve a small set (usually a few hundreds or thousands) of item candidates: $\mathcal{C}=\{v_i\}_{i=1}^{m}$, which might be relevant to $u$. The goal of the ranker is to score each candidate item $v$ in $\mathcal{C}$ and return the (top-$k$) ordered list as recommendations.  

\begin{figure}[t!]
	\centering
	\includegraphics[width=8.6cm]{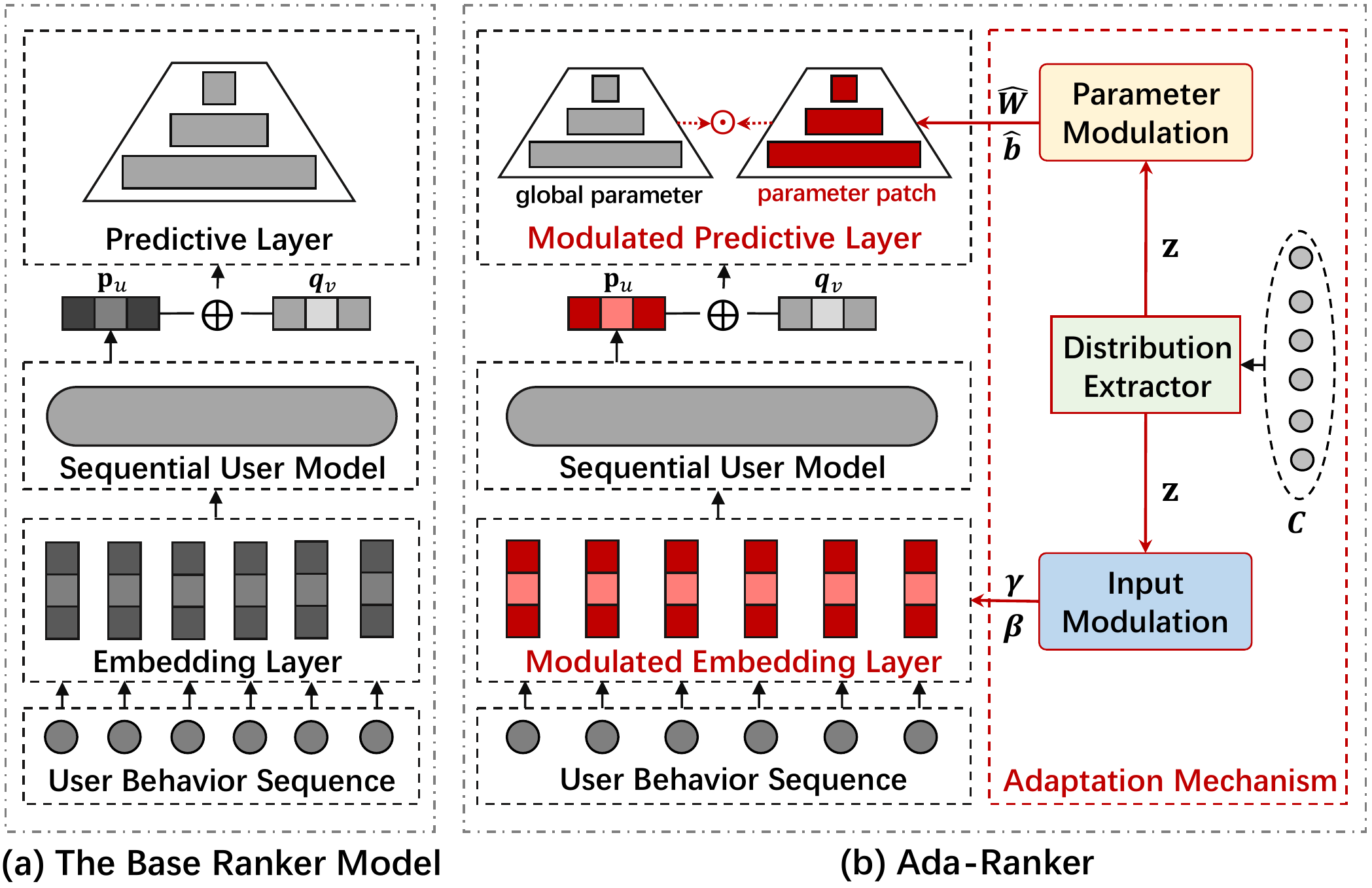}
	\vspace{-0.15in}
	\caption{An overview of the traditional sequential model (a) and Ada-Ranker paradigm (b). We use colored elements to indicate the new components in Ada-Ranker. }
	\label{fig:model} 
\end{figure}

\paratitle{A Standard Sequential Ranker Architecture}. For personalized recommendations, user $u$ is associated with a profiling representation, denoted by $\mathbf{x}_u$, which is derived based on 
her historical activities: $\mathbf{x}_u=\{v_1, v_2, ..., v_n\}$, where $n$ is the length of behaviors and $v_i$ is in chronological order. 
The ranker adopts a model $f$ to predict the preference score of user $u$ over the target candidate $v$: 
$\hat{y}_{uv}=f(\mathbf{x}_u, v)$, where there are usually three major layers in $f$: an embedding lookup layer $Q(\cdot)$, a sequential encoding layer $g^{SE}(\cdot)$, and a predictive layer $g^{PRED}(\cdot)$. 
We present a typical architecture of ranker models $f$ in Figure~\ref{fig:model}(a). The user behavior sequence $\{v_1, v_2, ..., v_n\}$ will first go through the embedding-lookup layer $Q(\cdot)$ to form the corresponding item embeddings:
\begin{equation}\label{eq-emb}
Q_u={Q}(\mathbf{x}_u)=\{\mathbf{q}_{v_1}, \mathbf{q}_{v_2}, ..., \mathbf{q}_{v_n}\}.
\end{equation}
Next, the sequential user encoder $g^{SE}(\cdot)$ will encode the item sequence and produce an embedding vector as user representation:
\begin{equation}\label{eq-SE}
\mathbf{p}_u=g^{SE}(Q_u).
\end{equation}
Then, $\mathbf{p}_u$ will be concatenated with a target item vector $\mathbf{q}_v$ (if there are some attribute features such as item category, user profiles, they can also be appended here) as the input to the predictive layer $g^{PRED}(\cdot)$, which is usually implemented as a simple two-layer Multi layer Perceptron (MLP) (see Figure~\ref{fig:parameter} (a)): 
\begin{equation}\label{eq-MLP}
\hat{y}_{uv} = g^{PRED}(\mathbf{p}_u, \mathbf{q}_v) = MLP(\mathbf{p}_u, \mathbf{q}_v).
\end{equation}
In summary, a standard ranker model $f$ can be instantiated as 
\begin{equation}
    \hat{y}_{uv} = f(\mathbf{x}_u, v) = g^{PRED}(g^{SE}(Q(\mathbf{x}_u)), \mathbf{q}_{v}).
\end{equation}

\paratitle{Potential Issues}. However, as introduced in Section~\ref{sec:intro}, the retrieved item candidates $\mathcal{C}$ may have different data distributions from diverse recall requests. Existing methods adopt a global ranker $f$ to serve all requests, and for a given user, it will produce the same score for item $v$, regardless of which candidate set $\mathcal{C}$ it draws from.
We argue that a better paradigm is  to leverage $\mathcal{C}$ as a ranking context, and let the model adjust itself according to the specific context to make more fine-grained and accurate prediction scores for the current ranking task. 

\section{Methodology}
\subsection{An Adaptive Ranking Paradigm}\label{sec:ranking_paradigm}
The essence of adaptive ranking paradigm is to incorporate a specially-designed adaptation mechanism  (aka, adaptor), which encodes the data distribution patterns of $\mathcal{C}$ and revises the global model $f$ to a local model $f'$ accordingly, so that $f'$ has a better discriminative capacity of ranking items in $\mathcal{C}$. 
To implement this new ranking paradigm, it is important to design the adaptation mechanism with special considerations for industrial applications: 
\begin{itemize}[leftmargin=*]
    \item {(C1) \emph{Lightweight}}. The model should be efficient in both computation and memory usage.
    \item {(C2) \emph{Model-agnostic}}. The model can be generalized to various sequential recommendation models, such as GRU-based~\cite{DBLP:journals/corr/HidasiKBT15}, CNN-based~\cite{10.1145/3289600.3290975} or GNN-based methods~\cite{DBLP:conf/aaai/WuT0WXT19}.
    \item {(C3) \emph{Plug-and-play}}. The adaptation mechanism can be applied as a model patch: the global model remains unchanged, so when the adaptor is disabled, the system performs exactly as how it works as usual; when the adaptor is enabled, the global model together with the model patch serve as the adaptation model and produce more accurate results. This feature ensures the flexibility to turn on/off the adaptor for different purposes.
\end{itemize}
To this end, we introduce a new ranking paradigm \textsl{Ada-Ranker}.
We first learn the distribution patterns $\mathbf{z}$ from $\mathcal{C}$, which will be used to modulate the global model $f$ and derive a local model $f'$; then, we use $f'$ to score the candidates in $\mathcal{C}$: $\hat{y}_{uv}=f'(\mathbf{x}_u, v)$. 
Since there are two key components in the global model $f$, i.e., $g^{SE}(\cdot)$ (Eq.~\ref{eq-SE}) and $g^{PRED}(\cdot)$ (Eq.~\ref{eq-MLP}), we propose two adaptation components, called \textsl{input modulation} and \textsl{parameter modulation}, which can incorporate the data distribution 
$\mathbf{z}$ to  modulate $g^{SE}(\cdot)$ and $g^{PRED}(\cdot)$ respectively. 
An overview of Ada-Ranker is illustrated in Figure~\ref{fig:model}(b) and in Algorithm~\ref{alg:Framwork}. In the next section, we will introduce the details of each proposed component.

\begin{figure}[t!]
	\centering
	\includegraphics[width=8cm]{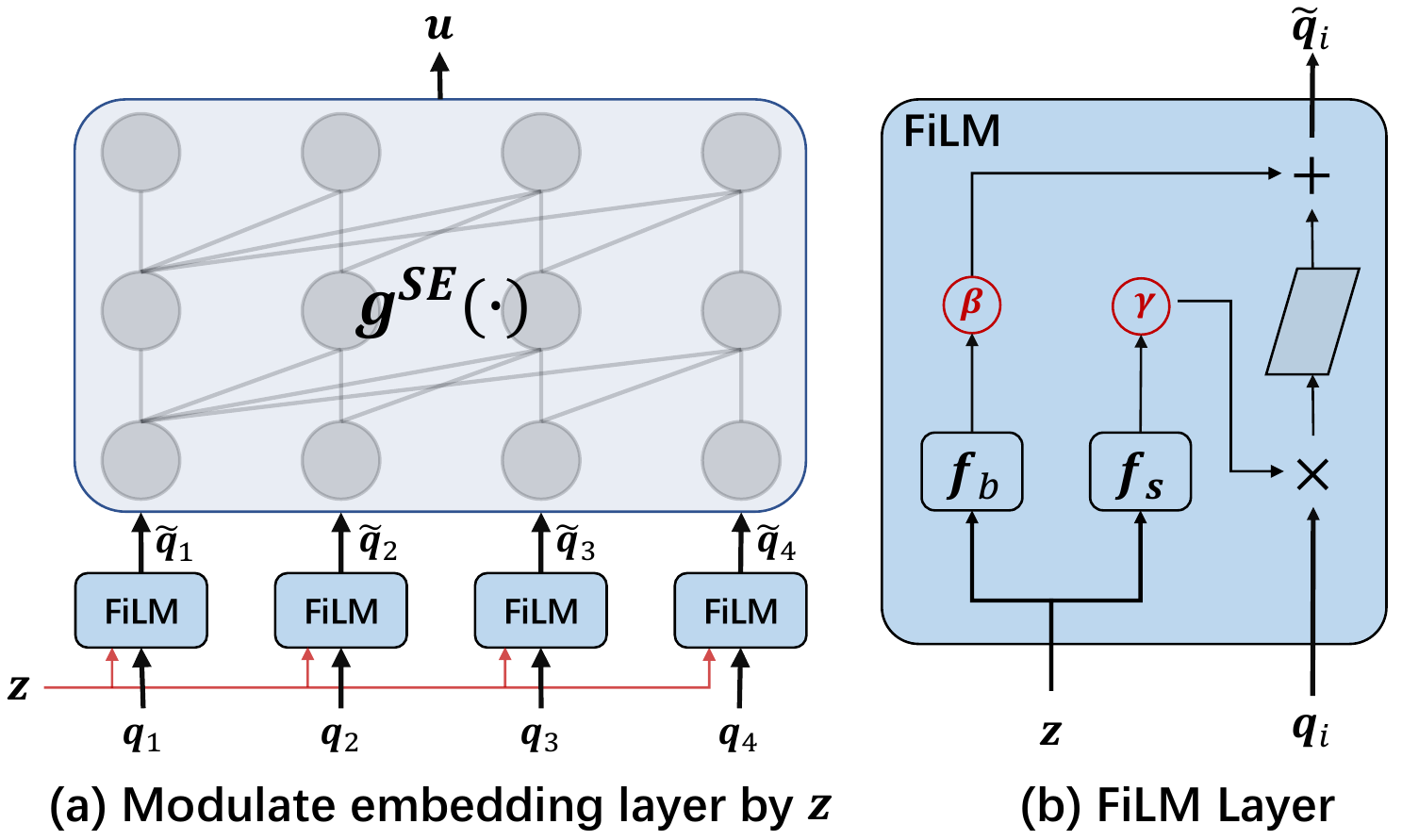}
	\caption{An illustration of the input modulation.}
	\label{fig:input} 
\end{figure}

\subsection{Data Distribution Learning from the Ranking Candidates} \label{sec:bias_ext}
To implement the ranking paradigm, the first issue is  how to effectively learn the data distribution patterns in ranking candidates $\mathcal{C}$, which extracts specific data characteristics for adjusting the ranker.

\paratitle{Neural Processes Encoder}. We assume that item candidates in $\mathcal{C}$ are drawn from a particular instantiation of stochastic process $\mathcal{F}$, which corresponds to a specific ranking request.
In order to characterize the dynamic and diverse data distributions, we borrow the idea of \emph{Neural Processes} (NP) from \cite{garnelo2018neural,garnelo2018conditional} to approximate a stochastic process with learnable neural networks. 
Recently, Neural Processes has been utilized for alleviating the user cold-start problem in recommender systems~\cite{10.1145/3442381.3449908}. 
Different from them, we adopt Neural Processes to model $\mathcal{F}$ which represents the data distribution associated with a ranking request. 
The fundamental advantages of NP lie in (1) providing an effective way to model data distributions conditioned on representations of observed data; and (2) being parameterized by Multi-layer Perceptron~(MLPs) which is more efficient than Gaussian Process (see more details in \cite{garnelo2018neural,garnelo2018conditional,10.1145/3442381.3449908}). Here we introduce the variational approximation implementation of NP encoder~\cite{10.1145/3442381.3449908}.

Specifically, we first generate a latent embedding vector $\mathbf{r}_j$ for each item $j$ in  $\mathcal{C}$ with a two-layer MLP:
\begin{equation}\label{eq:item_proj01}
	\mathbf{r}_j = MLP^{(NP)}(\mathbf{q}_j).
\end{equation}
Then, the NP encoder will aggregate these latent vectors to generate a permutation-invariant representation $\mathbf{r}$ via mean pooling:
\begin{equation}\label{eq:item_mean_pool}
	\mathbf{r} = (\mathbf{r}_1 + \mathbf{r}_2 + \dots \mathbf{r}_m)/m.
\end{equation}


\paratitle{Reparameterization}.  The above representation $\mathbf{r}$ will be used to generate the mean vector and the variance vector:
\begin{equation}
	\mathbf{s} = ReLU(\mathbf{W}_s\mathbf{r}),
\end{equation}
\begin{equation}
	\boldsymbol{\mu} = \mathbf{W}_\mu \mathbf{s}, \log\boldsymbol{\sigma} = \mathbf{W}_\sigma \mathbf{s}.
\end{equation}
Finally, the data distribution is modeled by a random variable $\mathbf{z}$: $\mathbf{z} \sim \mathcal{N}(\mathbf{\mu},\,\mathbf{\sigma}^{2})$, which is implemented with the reparameterization trick~\cite{DBLP:journals/corr/KingmaW13}, so that via making randomness in the input node of model, all the computational nodes in the model are differentiable and gradients can be smoothly backpropagated:
\begin{equation}\label{eq:repara}
	\mathbf{z} = \boldsymbol{\mu} + \boldsymbol{\epsilon} \odot \boldsymbol{\sigma}, \ \ \boldsymbol{\epsilon} \sim \mathcal{N}(0,\,\mathbf{I}),
\end{equation}
where $\odot$ means the element-wise product operation.

After extracting the data distribution from the candidates, we next discuss how to adapt a ranker to the given data distribution in two aspects, namely \emph{input modulation} and \emph{parameter modulation}. 

\subsection{Input Modulation}
To ensure the model-agnostic property, we modulate the latent representations of input sequence, and this operation can be shared by different sequential recommendation models.

\paratitle{Modeling Data Distributions as Adaptation Conditions}. To revise the item representations according to the underlying data distributions, our idea is to consider the extracted $\mathbf{z}$ in Eq.~\ref{eq:repara} as a condition to adjust the corresponding input representations. Here, we adopt the FiLM method~\cite{DBLP:conf/aaai/PerezSVDC18} that is a general-purpose method to model the influence of conditioning information on neural networks.
 Conditioned on the distribution patterns $\mathbf{z}$, we perform linear modulation on the latent representations of item sequence, such that input embeddings are adjusted to new representations that are more distinguishable among candidate set $\mathcal{C}$. The details of this adaptation process for input modulation are shown in Figure~\ref{fig:input}.

Specifically, We learn to generate two modulation coefficients $\gamma$ and $\beta$  through the conditional representation $\mathbf{z}$:
\begin{equation}
	\label{eq:im_1}
	\gamma = f_s(\mathbf{z}), \ \  \beta = f_b(\mathbf{z}),
\end{equation}
where $f_s$ and $f_b$ are two neural networks with different parameters, formulated as: $f(\mathbf{z}) = \mathbf{w}_2ReLU(\mathbf{W}_1 \mathbf{z} + \mathbf{b}_1) + b_2$. The latent representations of item sequence are then adjusted by:
\begin{equation}
	\label{eq:im_2}
	\tilde{\mathbf{q}}_t = \gamma \mathbf{q}_t + \beta.
\end{equation}
Here,  $f_s$ and $f_b$ are shared for all items in behavior sequence.

\paratitle{A Case with GRU-based Recommender}. For example, if the sequential model $g^{SE}(\cdot)$ is GRU, then the modulated user vector is: $\tilde{\mathbf{u}}=GRU(\{\tilde{\mathbf{q}}_t\}_{t=1}^n)$.
Ada-Ranker does not change any parameters in GRU (due to the model-agnostic property). Instead, it makes adaptations on the data input, which creates room for the encoded user vector to be more dedicated to a provided distribution $\mathbf{z}$.

\begin{figure}[t!]
	\centering
	\includegraphics[width=8.5cm]{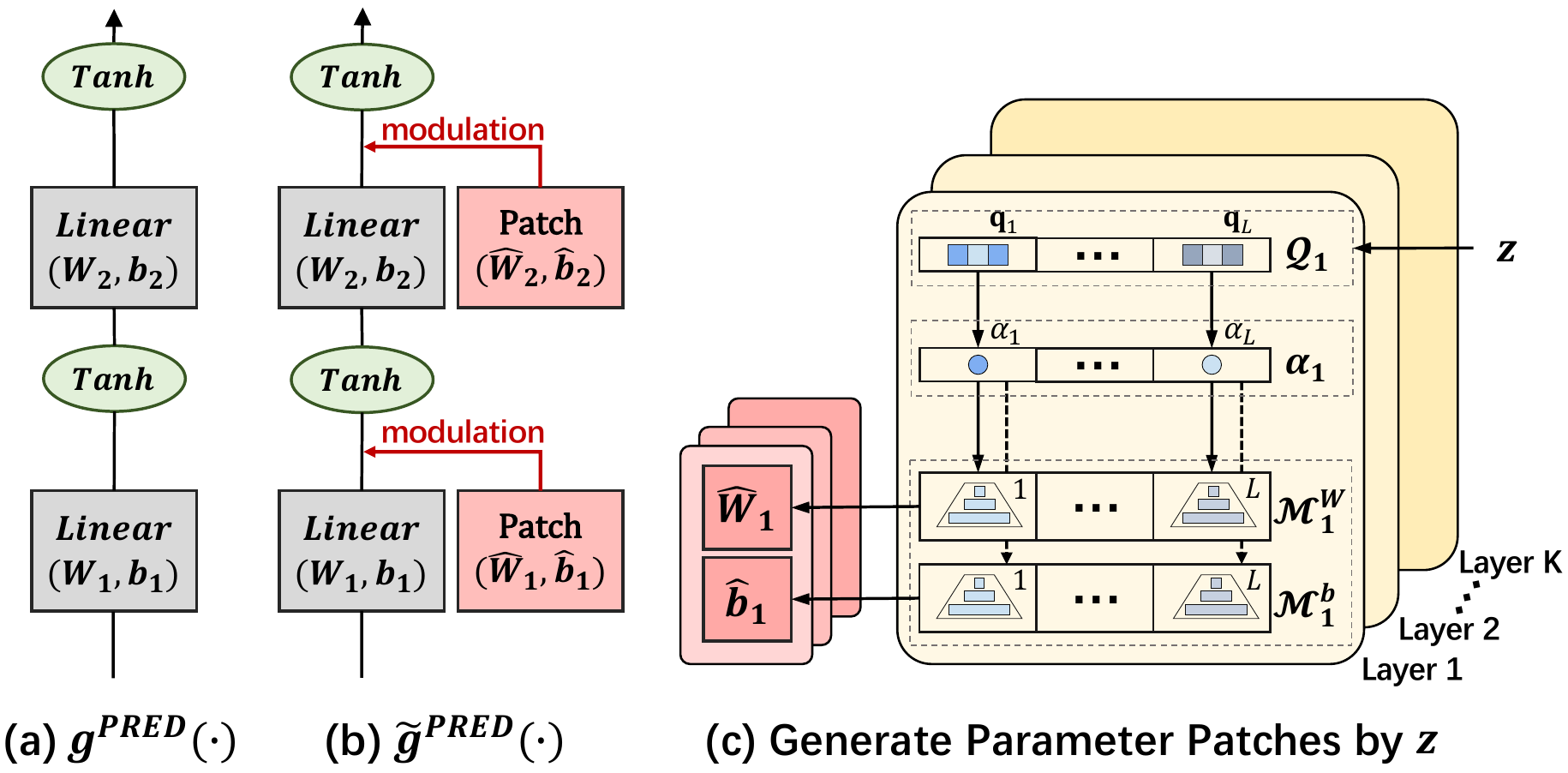}
	\caption{An illustration of the parameter modulation.
	}
	\label{fig:parameter} 
\end{figure}
\subsection{Parameter Modulation}
For the modulation of scoring function $g^{PRED}(\cdot)$, we adopt the idea of model patch~\cite{DBLP:conf/icml/HoulsbyGJMLGAG19,sheng2021model} to adjust the parameters, which is a light-weight approach to implement parameter adaptation. 


\paratitle{Adaptation by Model Patch}.
The basic idea of model patch is to learn parameter variations for adapting to new input or task context~\cite{10.1145/3397271.3401156}. 
As shown in Figure~\ref{fig:parameter} (a)-(b), we can incorporate another two MLPs to generate \emph{parameter patches} for the $k$-th hidden layer of the predictive layer according to the conditional representation $\mathbf{z}$ which represents the underlying data distributions:
\begin{equation}
\label{eq:pm_mlp}
    \widehat{\mathbf{W}}_k = MLP^{(w_k)}(\mathbf{z}), \ \  \widehat{\mathbf{b}}_k = MLP^{(b_k)}(\mathbf{z}),
\end{equation}
Then the ranker's MLP parameters (Eq.~\ref{eq-MLP}) are modulated by:
\begin{equation}
    \label{eq:pm_modulate}
     \widetilde{\mathbf{W}}_k = {\mathbf{W}}_k \odot \widehat{\mathbf{W}}_k, \ \ \widetilde{\mathbf{b}}_k = {\mathbf{b}}_k + \widehat{\mathbf{b}}_k,
\end{equation}
where $\odot$ denotes the element-wise multiplication and $k$ means the $k$-th layer of the ranker's MLP. Let $h$ denote the input dimension and $d$ denote the output dimension, we have $\mathbf{W}_k, \widehat{\mathbf{W}}_k, \widetilde{\mathbf{W}}_k \in \mathbb{R}^{h \times d}$ and $\mathbf{b}_k, \widehat{\mathbf{b}}_k, \widetilde{\mathbf{b}}_k \in \mathbb{R}^{d}$. We will replace the original MLPs' parameters ${\mathbf{W}}_k$ by $\widetilde{\mathbf{W}}_k$ and ${\mathbf{b}}_k$ by $\widetilde{\mathbf{b}}_k$. In this way, a global scoring function $g^{PRED}(\cdot)$ is adapted into a local scoring function $\Tilde{g}^{PRED}(\cdot)$.

\paratitle{Learning with Parameter Pool}. A major problem of  Eq.~(\ref{eq:pm_mlp}) is that the modulation parameters are generated in 
an unconstrained way, and it has been shown that it is difficult to directly optimize MLPs to approximate  arbitrary real-valued vectors, especially when the training sample is not sufficient~\cite{krizhevsky2012imagenet}.
Thus, we propose to construct a parameter pool consisting of multiple base parameters (either vector or matrix), which is a parameter memory network~\cite{DBLP:conf/icml/SantoroBBWL16}. Furthermore, the parameters are derived based on a linear combination of multiple base parameters. The proposed adaptation mechanism for parameter modulation is depicted in Figure~\ref{fig:parameter} (c).
Specifically,  we have a  parameter pool $\mathcal{M}$  with $L$ base parameters: $\mathcal{M}=\{ \mathbf{M}_1, \mathbf{M}_2, ..., \mathbf{M}_L \}$, where each $\mathbf{M}_i$ has the same shape with the MLP parameter ${\mathbf{W}}$ of the ranker. Then, the parameter patch $\widehat{\mathbf{W}}$ is composited by a linear combination of  $\mathcal{M}$ as follows:
\begin{equation}
\label{eq:para_mn}
   \widehat{\mathbf{W}} = \sum_{i=1}^{L} \alpha_i \mathbf{M}_i,
\end{equation}
The coefficients $\alpha_i$ are determined by a set of reading heads $\mathcal{Q}=\{ \mathbf{q}_1, \mathbf{q}_2, ..., \mathbf{q}_L \}$:
\begin{align}
    a_i &= \mathbf{z} \cdot \mathbf{q}_i \label{eq:get_zi},\\
    \alpha_i &=  \frac{\exp( a_i)}{\sum_{j=1}^{L} \exp(  a_j)},  \forall i=1,2,...,L \label{eq:alpha_softmax},
\end{align}
where $\mathbf{q}_*$ are learnable vectors which have the same shape with $\mathbf{z}$. We use individual parameter pools ${\mathcal{M}_k}, \mathcal{Q}_k$ for each layer of the MLP parameters in $g^{PRED}(\cdot)$ (e.g., $\mathbf{W}_k$ in the $k$-th layer). The case of the bias parameters $\mathbf{b}_*$ is the same, which is omitted here.  


\subsection{Optimization and Discussion} \label{sec:parameter}

After the input modulation and parameter modulation, the ranker transforms from a global model $f(u,v)=g^{PRED}(g^{SE}(Q(\mathbf{x}_u)), \mathbf{q}_{v})$ to a local model $f'(u,v)=\Tilde{g}^{PRED}(\Tilde{g}^{SE}(Q(\mathbf{x}_u)), \mathbf{q}_{v})$.
We next discuss how to optimize the proposed ranking approach.

\paratitle{Parameter Learning}. 
The optimization method remains the same as the original ranker, by minimizing the binary cross entropy loss:
\begin{equation}
    \label{eq:loss}
    \mathcal{L} =\frac{1}{|\Gamma|}\sum_{(u, v)\in \Gamma}-y_{u,v} \log \Tilde{y}_{uv} - (1-y_{u,v})\log(1-\Tilde{y}_{uv}),
\end{equation}
where $\Tilde{y}_{uv}=f'(u,v)$ and $\Gamma$ denotes the set of training instances.
There are several feasible strategies to train the Ada-Ranker. For clarity, let $\Theta$ denote the set of parameters of the base model $f$, and $\Phi$ denote the set of parameters in adaptation modules (including the distribution extractor, input modulation and parameter modulation).
We have the following three alternative training strategies:
 
\begin{enumerate}
 [leftmargin=*]
	\item $\emptyset \Rightarrow \Theta + \Phi$. A straightforward approach is to train the whole Ada-Ranker model in an end-to-end manner from scratch ($\emptyset$).
	\item $\Theta \Rightarrow \Theta + \Phi$. In the first stage, we train the base model until convergence, so only $\Theta$ get updated in this stage; in the second stage, we load the pre-trained base model and train the whole Ada-Ranker, so parameters in both $\Theta$ and $\Phi$ will be updated.
	\item $\Theta \Rightarrow  \Phi$. In the first stage, we train the base model until it converges, so only $\Theta$ get updated in this stage; in the second stage, we load the pre-trained base model $f$ and freeze its parameters, and train the Ada-Ranker by only updating  $\Phi$.
\end{enumerate} 
 
To achieve the (C3) property mentioned in Section~\ref{sec:ranking_paradigm}, we adopt the last training strategy. Pre-training the base model's parameters can offer a good parameter initialization for the training of Ada-Ranker, so the first strategy,  $\emptyset \Rightarrow \Theta + \Phi$, is usually worse than the other two strategies. The major merit of the third strategy is that, after Ada-Ranker's training, the base model's parameters remain unchanged, such that an industrial system can flexibly turn on or turn off the adaptation module to meet diverse requirements of different scenarios.
The overall training process of Ada-Ranker is summarized in Algorithm~\ref{alg:Framwork} (see experimental comparisons of the training strategies in Section~\ref{sec:traing_method}). For inference stage, we only need to run Line \ref{algo:extract} to Line \ref{algo:score} in Algorithm~\ref{alg:Framwork} with a trained  Ada-Ranker model.

\begin{algorithm}
	\caption{The training procedure of Ada-Ranker.}
	\label{alg:Framwork}
	\small
	\begin{algorithmic}[1]
		\Require
		Training instances: a set of \{behavior sequence $\mathbf{x}_u=\{v_1, v_2, ..., v_n\}$,  labeled items $\mathcal{C}=\{v_i, y_{uv_i}\}_{i=1}^{m}$\}.
		\Ensure
		An Ada-Ranker model $f'(\mathbf{x}_u, v)$.
		\State  {// The first training stage}
		\State Train base model $f(\mathbf{x}_u, v)$ and obtain pre-trained global parameters $\Theta$.
		\State  {// The second training stage}
		\State Initialize base model parameters in Ada-Ranker $f'(\mathbf{x}_u, v)$ by $\Theta$, initialize adaptation parameters $\Phi$ randomly.
		
		\While{not convergence}
		\State Fetch a training instance \{$\mathbf{x}_u$, $\mathcal{C}$\}
		\State  {// Distribution Extraction}
		\State Extract the distribution pattern $\mathbf{z}$ by Eq.(\ref{eq:item_proj01} - \ref{eq:repara}). \label{algo:extract}
		\State  {// Input Modulation}
		\State Generate $\gamma, \beta$ with $\mathbf{z}$ by Eq.(\ref{eq:im_1}).
		\State Use $\gamma, \beta$ to modulate each $\mathbf{q}_t$ by Eq.(\ref{eq:im_2}) to get $\tilde{\mathbf{q}}_t$.
		\State Encode $\{\tilde{\mathbf{q}}_t\}_{t=1}^n$ and generate $\tilde{\mathbf{u}}$ by $g^{SE}(\cdot)$.
		\State  {// Parameter Modulation}
		\State Calculate the coefficients $\mathbf{\alpha}$ by $\mathcal{Q}_k$ and $\mathbf{z}$ in Eq.(\ref{eq:get_zi})\&(\ref{eq:alpha_softmax}).
		\State Generate parameters $\widehat{\mathbf{W}}_k$, $\widehat{\mathbf{b}}_k$ in Eq.(\ref{eq:para_mn}).
		\State Modulate $g^{PRED}$ parameters $\mathbf{W}_k, \mathbf{b}_k$ in Eq.(\ref{eq:pm_modulate}).
		\State Use the revised $\Tilde{g}^{PRED}$ to score the candidates in $\mathcal{C}$. \label{algo:score}
		\State Calculate loss $\mathcal{L}$ in Eq.(\ref{eq:loss}).
		\State Stop gradients for $\Theta$ and update $\Phi$ by Adam optimizer.
		\EndWhile
	\end{algorithmic}
\end{algorithm}

\paratitle{Complexity Discussion}.
The number of additional parameters $\Phi$ is $O(LK (h  d+d) + d   d)$, while the original global model has parameter size $|\Theta|=O(M d + K (h  d+d) + |\theta|)$, where $L$, $K$, $h$, $d$ , $M$, $\theta$ denote the number of parameter pool slots, the number of hidden layers of the predictive layer, hidden layer size for neural networks, item embedding size, the number of items and the parameters of $g^{SE}(\cdot)$ respectively. For a recommender system, $M$ (usually thousands or millions) is  much larger than $L$ (e.g., 10 in this paper), $K$ (e.g., 2 in this paper) and $d$ (e.g., 64 in this paper), compared with other methods that need to train and maintain multiple versions of $\Theta$ for serving different scenarios, our Ada-Ranker is much more parameter economic. As for the computational cost, Neural Processes perform the calculation on candidate items individually (see Eq.(\ref{eq:item_proj01})), without the requirement of interactions among items in the candidate list. Thus, with distribution learning techniques, the computational cost is linear with the number of items in the candidate list. Besides, for the input modulation and parameter modulation, the forward calculation does not depend on a specific candidate item, so the whole calculation only needs to be conducted once. Thus, the overall computational cost can be effectively controlled at a reasonable level.
\begin{table}
	\small
	\caption{Statistics of datasets after preprocessing.}
	\vspace{-0.15in}
	\label{tab:datasets}
	\setlength{\tabcolsep}{1mm}{
		\begin{tabular}{lrrrcrc}
			\toprule
			Dataset  & \# Users & \# Items & \# Actions & \# Avg.len & Sparsity  & \#Categories \\
			\midrule
			ML10M 	& 69,878 & 10,676 & 10,000,047 & 143 & 98.66\%  & 18 \\
			Taobao	& 487,813 & 787,094 & 24,938,811 & 34 & 99.99\%  & 20\\
			Xbox 	& 1,000,000 & 6,071 & 25,315,983 & 15 & 99.58\%  & 12 \\
			\bottomrule
		\end{tabular}
	}
\end{table}

\begin{table*}[t!]
	\caption{Performance comparison of seven base models and Ada-Ranker on three datasets. }
	\label{tab:basemodel}
	\vspace{-0.1in}
	\setlength{\tabcolsep}{0.8mm}{
		\begin{tabular}{c|l|cc|cc|cc|cc|cc|cc|cc} 
			\toprule
			\multicolumn{1}{c|}{\multirow{2} * {Datasets}} &\multicolumn{1}{c|}{\multirow{2} * {Models}}&\multicolumn{2}{|c}{MF}&\multicolumn{2}{|c}{GRU4Rec}&\multicolumn{2}{|c}{SASRec}&\multicolumn{2}{|c}{NARM}&\multicolumn{2}{|c}{NextItNet}&\multicolumn{2}{|c}{SHAN}&\multicolumn{2}{|c}{SRGNN}\\
			
			&  & GAUC & NDCG & GAUC & NDCG & GAUC & NDCG & GAUC & NDCG & GAUC & NDCG & GAUC & NDCG & GAUC & NDCG\\
			\midrule
			\multirow{2} * {ML10M}
			& Base & 0.8683 & 0.6942 & 0.9187 & 0.7781 & 0.9106 & 0.7616 & 0.9156 & 0.7718 & 0.8620 & 0.6855 & 0.8667 & 0.6973 & 0.8993 & 0.7441 \\
			
			& Ada-Ranker & \textbf{0.8807} & \textbf{0.7121} & \textbf{0.9279} & \textbf{0.7932} & \textbf{0.9214} & \textbf{0.7783} & \textbf{0.9261} & \textbf{0.7879} & \textbf{0.8778} & \textbf{0.7069} & \textbf{0.8816} & \textbf{0.7164} & \textbf{0.9161} & \textbf{0.7676} \\
			\hline 
			
			\multirow{2} * {Taobao}
			& Base & 0.8363 & 0.6605 & 0.8734 & 0.7208 & 0.8707 & 0.7211 & 0.8707 & 0.7177 & 0.8482 & 0.6872 & 0.8609 & 0.7094 & 0.8637 & 0.7105 \\
			
			& Ada-Ranker & \textbf{0.8512} & \textbf{0.6907} & \textbf{0.8888} & \textbf{0.7490} & \textbf{0.8864} & \textbf{0.7481} & \textbf{0.8881} & \textbf{0.7439} & \textbf{0.8625} & \textbf{0.7079} & \textbf{0.8653} & \textbf{0.7106} & \textbf{0.8791} & \textbf{0.7322} \\
			\hline
			\multirow{2} * {Xbox}
			& Base & 0.9036 & 0.7676 & 0.9388 & 0.8302 & 0.9323 & 0.8172 & 0.9368 & 0.8258 & 0.9343 & 0.8222 & 0.9217 & 0.8022 & 0.9336 & 0.8212 \\
			
			& Ada-Ranker & \textbf{0.9123} & \textbf{0.7897} & \textbf{0.9451} & \textbf{0.8438} & \textbf{0.9382} & \textbf{0.8291} & \textbf{0.9412} & \textbf{0.8354} & \textbf{0.9401} & \textbf{0.8336} & \textbf{0.9282} & \textbf{0.8131} & \textbf{0.9426} & \textbf{0.8384} \\
			
			\bottomrule
		\end{tabular}
	}
\end{table*}

\section{Experiment}
\subsection{Experimental Settings}  
\subsubsection{Datasets}
\label{sec:setting01}
We use three datasets for experiments, covering movie, e-commerce  and gaming recommendation scenarios. 
\begin{itemize}[leftmargin=*]
	\item[$\bullet$] \textbf{ML10M Dataset}~\footnote{\url{https://grouplens.org/datasets/movielens/10m/}} is a widely used benchmark dataset. It contains 10,000,054 ratings and 95,580 tags applied to 10,681 movies by 71,567 users from an online movie recommender service.
	There are in total 18 categories (genres in origin file) of items (with overlapping items): \textsl{Action}, \textsl{Adventure}, \textsl{Animation}, \textsl{Children}, \textsl{Comedy}, \textsl{Crime}, \textsl{Documentary}, \textsl{Drama}, \textsl{Fantasy}, \textsl{Film-Noir}, \textsl{Horror}, \textsl{Musical}, \textsl{Mystery}, \textsl{Romance}, \textsl{Sci-Fi}, \textsl{Thriller}, \textsl{War} and \textsl{Western}.
	
	\item[$\bullet$] \textbf{Taobao Dataset}~\footnote{\url{https://tianchi.aliyun.com/dataset/dataDetail?dataId=649&userId=1}} is an e-commerce dataset released by Taobao. The user-item interactions logs include the type of interaction (e.g., purchase or click), timestamps and items' categories (in anonymous).  The item category distribution is severely uneven, so we keep the top 19 categories which contain most items, and regard the remaining items belong to a virtual category \textsl{others}.
	
	\item[$\bullet$] \textbf{Xbox Dataset} is a private dataset provided by Xbox, an online gaming platform of Microsoft. It contains one year's user-game playing logs spanning from September 2020 to September 2021. We sample 1 million users from the original dataset. There are 12 major categories of items in Xbox dataset (with overlapping items), such as \textsl{GamePass}~\footnote{GamePass is membership subscription, with which users can play a library of games on Xbox by paying a certain amount of money each month.}, \textsl{Action Adventure} and \textsl{Card Board}.
	
\end{itemize}

For all datasets, we remove users with fewer than ten interaction records, group the interaction records by users and sort them in chronological order. The basic statistics are reported in Table~\ref{tab:datasets}. Following previous works~\cite{DBLP:conf/icdm/KangM18, DBLP:conf/cikm/ZhouWZZWZWW20}, we apply the \textit{leave-one-out} strategy for evaluation. For each user interaction sequence, the last item
and the item before it are considered as test and validation data, respectively, and the remaining items is for training. 
A group of candidate items contains 1 positive instance and 19 negative samples.

Since the focus of this paper is to empower rankers with adaptation ability for dynamic ranking context, to verify this property, the key part of dataset preparation lies in the negative sampling strategy, which reflects that the ranker needs to discriminate positive instance from different groups of item candidates. For example, if instances are sampled by item popularity, we can simulate the situation that the user is visiting the \textsl{Most Popular} tab; if instances are all sampled from the \textsl{Shooter} category, it indicates that a user has clicked into the \textsl{Shooter} channel and searches for games in this scope. Thus, to make the test environment more comprehensive and the distribution of candidates more diverse, we sample negative items for each positive instance by the \textsl{distribution-mixer sampling}.

\textbf{\textit{Distribution-mixer sampling}}: We consider two basic factors that influence the distribution of candidates, namely
(1) item categories and (2) item popularity. 
Our negative sampling strategies are designed as:
(Step-1) draw a random number $d$ from $\{1,2,3\}$, indicating that $d$ item categories will be involved in the 19 negative instances. If $d>1$, then randomly sample $d-1$ item categories which are different with the positive item's category;
(Step-2) conduct a Bernoulli trial with $p=0.5$, if the outcome is \textsl{success}, set the item sampling strategy to \textsl{popularity-bias}, otherwise set the item sampling strategy to \textsl{uniform} for this round;
(Step-3) from each $d$ sampled categories, sample items with strategy returned in Step-2, so that the total number of items is 19. Step-1 to Step-3 is repeated individually for each positive instance in the train/valid/test set.


\subsubsection{Evaluation Metrics}
Following the previous works~\cite{RecBole, Evaluating_TopN}, we adopt two widely used metrics for evaluation the ranking results: GAUC (Area Under the ROC curve, group by each user), and NDCG (Normalized Discounted Cumulative Gain).  

\subsubsection{Implementation Details}
All methods are implemented with Python 3.8 and PyTorch 1.8.1. We use four Linux servers with the same hardware configuration: CPU is \textsl{Intel(R) Xeon(R) E5-2690 v4 @ 2.60GHz} and GPU is \textsl{Tesla P100}.  
We use the Adam optimizer with a learning rate of 0.001, where the batch size is set as 4096. The dropout rate is set to 0.4. The dimension of the embedding is 64, and the hidden states' size in the prediction layer is $128\ast64$ and $64\ast1$. The maximum sequence length is 200, 100 and 50 for ML10M, Taobao and Xbox datasets, respectively.
For Ada-Ranker, the slots of memory network $L$ is set to 10. The source code of Ada-Ranker is released at \url{https://github.com/RUCAIBox/Ada-Ranker}.

\subsection{Improving Various Base Models}
\subsubsection{Base Models} \label{sec:baseline}
Ada-Ranker is a model-agnostic ranking paradigm which can enhance various types of sequential recommendation models. To verify this, we test Ada-Ranker's performance based on the following representative sequential models: 
\begin{itemize}[leftmargin=*]
	\item[$\bullet$] \textbf{MF}~\cite{DBLP:journals/computer/KorenBV09} factorizes each item and user into an embedding vector based on the user-item interaction logs. It is the only one base model that is non-sequential.
	\item[$\bullet$] \textbf{GRU4Rec}~\cite{DBLP:journals/corr/HidasiKBT15} uses GRU to model a user's behavior sequence. It is one of the most popular model for sequential recommendations.
	\item[$\bullet$] \textbf{SASRec}~\cite{DBLP:conf/icdm/KangM18} uses the unidirectional multi-head self-attention model to deal with a user's behavior sequence. 
	\item[$\bullet$] \textbf{NARM}~\cite{DBLP:conf/cikm/LiRCRLM17} is a session-based recommendation approach with RNN. It uses an attention mechanism to determine the relatedness of the past purchases in the session for the next purchase.
	\item[$\bullet$] \textbf{NextItNet}~\cite{10.1145/3289600.3290975} applies dilated CNN for long sequence modeling.
	\item[$\bullet$] \textbf{SHAN}~\cite{DBLP:conf/ijcai/YingZZLXXX018} is a 2-layer hierarchical attention network to model user's dynamic long-term preference and sequential behaviors.
	\item[$\bullet$] \textbf{SRGNN}~\cite{DBLP:conf/aaai/WuT0WXT19} models session sequences as graph structured data and uses a GNN to capture complex transitions of items.
\end{itemize}

\begin{table*}[t!]
	\caption{Performance comparison with eight different baselines based on two base models (GRU4Rec and SASRec).  The best performance and the second best performance methods are denoted in bold and underlined, respectively. The $p$-value of significance test is performed in the NDCG metrics of Ada-Ranker with the corresponding best baseline method (underlined).}
	\label{tab:baselines}
	\vspace{-0.15in}
	\setlength{\tabcolsep}{1.6mm}{
		\begin{tabular}{l|cc|cc|cc|cc|cc|cc} 
			\toprule
			\multicolumn{1}{c|}{\multirow{3} * {Methods}} &\multicolumn{6}{|c|}{GRU4Rec as base model}&\multicolumn{6}{|c}{SASRec as base model}\\
			\cline{2-13}
			& \multicolumn{2}{|c|}{ML10M} & \multicolumn{2}{|c|}{Taobao} & \multicolumn{2}{|c|}{Xbox}  & \multicolumn{2}{|c|}{ML10M} & \multicolumn{2}{|c|}{Taobao} & \multicolumn{2}{|c}{Xbox}\\
			
			& GAUC & NDCG & GAUC & NDCG & GAUC & NDCG & GAUC & NDCG & GAUC & NDCG & GAUC & NDCG \\
			\midrule
			Base & 0.9187 & 0.7781 & 0.8734 & 0.7208 & 0.9388 & 0.8302 & 0.9106 & 0.7616 &  0.8707 & 0.7211 & 0.9323 & 0.8172  \\
			
			DNS & 0.9005 & 0.7512 & 0.8646 & 0.7200 & 0.9254 & 0.8182 & 0.8942 & 0.7411 & 0.8632 & 0.7183 & 0.9219 & 0.8125 \\
			
			GSF$_{concat}$ & 0.9150 & 0.7744 & \underline{0.8903} & \underline{0.7433} & 0.9335 & 0.8261 & 0.9072 & 0.7623 & 0.8858 & \underline{0.7400} & 0.9308 & 0.8201 \\
			
			GSF$_{avg}$ & 0.9205 & 0.7760 & 0.8860 & 0.7370 & 0.9390 & 0.8325 & 0.9116 & 0.7622 & 0.8798 & 0.7314 & 0.9353 & 0.8270 \\
			
			GSF$_{trm}$ & 0.9232 &  0.7815 & 0.8865 & 0.7366 & 0.9429 & \underline{0.8414} & 0.9161 & 0.7700 & 0.8836 & 0.7351 & 0.9366 & 0.8278 \\
			
			PD & \underline{0.9254} & 0.7818 & \textbf{0.8913} & 0.7387 & 0.9373 & 0.8216 & \underline{0.9196} & 0.7705 & \textbf{0.8878} & 0.7323 & 0.9325 & 0.8137  \\
			
			DecRS & 0.9247 & \underline{0.7881} &  0.8785 & 0.7306 & 0.9427 & 0.8367 & \underline{0.9195} & 0.7699 & 0.8776 & 0.7299 & 0.9345 & 0.8227 \\
			
			DLCM & 0.9240 & 0.7851 &  0.8835 & 0.7366 & 0.9426 & 0.8374 & 0.9177 & 0.7713 & 0.8793 & 0.7347 & 0.9376 & 0.8272 \\
			
			PRM & 0.9244 & 0.7863 &  0.8829 & 0.7397 & \underline{0.9442} & 0.8396 & 0.9181 & \underline{0.7731} & 0.8788 & 0.7345 & \underline{0.9379} & \underline{0.8288} \\
			
			\hline
			AdaRanker & \textbf{0.9279} & \textbf{0.7932} &  0.8888 & \textbf{0.7490} & \textbf{0.9451} & \textbf{0.8438} & \textbf{0.9214} & \textbf{0.7783} & \underline{0.8864} & \textbf{0.7481} & \textbf{0.9382} & \textbf{0.8291} \\
			\hline
			$p$-value & \multicolumn{2}{|c|}{7.02e-07} & \multicolumn{2}{|c|}{7.64e-09} & \multicolumn{2}{|c|}{0.00424}  & \multicolumn{2}{|c|}{0.000215} & \multicolumn{2}{|c|}{5.39e-12} & \multicolumn{2}{|c}{0.1586} \\
			\bottomrule
		\end{tabular}
	}
\end{table*}

\subsubsection{Results}
We train all base sequential models and their corresponding Ada-Rankers on three datasets. The results are reported in Table~\ref{tab:basemodel}. 
We observe that Ada-Ranker outperforms all base sequential models substantially and consistently, on all datasets, and in terms of overall metrics. Specifically, the average improvement of Ada-Ranker~(GRU4Rec) over GRU4Rec on the three datasets is 2.50\% in terms of NDCG and the improvement of Ada-Ranker~(SASRec) over SASRec is 2.46\%. Since Ada-Ranker and its base model have the same embedding layer, user model architectures and prediction layers, it shows that our proposed Ada-Ranker is effective to adapt a base mode according to different distributions of candidates for the current task.
Note that we adopt the $\Theta \Rightarrow  \Phi$ strategy (see Section \ref{sec:parameter}) to train Ada-Ranker, which first trains the base model and then only fine-tune parameters of adaptation networks. The impressive performance gain demonstrates the effectiveness of our adaptation mechanism, even though we just fine-tune the small set of parameters of adaptation networks. In summary, Ada-Ranker processes the virtue of the plug-and-play property: enhancing any given base model with adaptation modules while maintaining the original base model unchanged.

\subsection{Comparison with Baselines}\label{sec:baselines}

\subsubsection{Baselines}
To the best of our knowledge, there are no studies on dynamic model adaptation for rankers in literature. However, we can extend some existing methods to make them able to handle dynamic candidate sets, so that they can serve as baselines. Generally, we collect baselines from four perspectives and adopt GRU4Rec and SASRec as base models to test how different types of methods perform.

\noindent (1) From the view of dynamic negative sampling:

$\bullet$ \textbf{DNS}~\cite{DBLP:conf/sigir/ZhangCWY13} dynamically chooses negative samples from the ranked list produced by the current prediction model and iteratively update it. 
Considering that DNS dynamically constructs candidate list in the training stage, it is useful to adapt the base model to various types of data distribution.  

\noindent (2) From the view of group-wise ranking:

$\bullet$ \textbf{GSF}~\cite{DBLP:conf/ictir/AiWBGBN19} uses a groupwise scoring function, in which the relevance score of an item is determined jointly by all items in the candidate list. We implement three types of scoring functions based on the idea of GSF. (1) \textsl{\underline{GSF$_{concat}$}}: concatenates all candidates' embeddings and produces scores for all candidates, which is the original implementation of {GSF}~\cite{DBLP:conf/ictir/AiWBGBN19}.  (2) \textsl{\underline{GSF$_{avg}$}}: averages all candidates' embeddings into one embedding vector (which can be regarded as a context vector), and appends it to the input of scoring function.  (3) \textsl{\underline{GSF$_{trm}$}}: instead of simply using averaging, it uses a 2-layer Transformer to encoding the candidate list and get all items' contextualized representation, then append the corresponding contextualized representation to the input of scoring function.

\noindent (3) From the view of causal analysis:

$\bullet$ \textbf{PD}~\cite{ZhangF0WSL021} removes the confounding popularity bias in model training and adjusts the recommendation score with desired popularity bias via causal intervention. The major consideration is to decouple the user-item matching with item popularity: $f_{\theta}(u,i,m_i^t)=f_{\theta}(u,i) \cdot g(m_i^t)$, where $m_i^t$ is the item $i$’s present popularity calculated by a statistical method. We can draw an analogy that the popularity bias in PD corresponds to the data distribution bias in our scenario. Thus, in our experiment, we replace $m_i^t$ with our distribution vector and use a $MLP(\cdot)$ to calculate the refining score.

$\bullet$ \textbf{DecRS}~\cite{DBLP:conf/kdd/WangF0WC21} explicitly models the causal relations of user representations during training, and leverages backdoor adjustment to eliminate the impact of the confounder which causes the bias amplification. In our experiment, we replace the group-level user representation in original model with a category-level representation of the candidate list. 

\noindent (4) From the view of re-ranking:

$\bullet$ \textbf{DLCM}~\cite{AiBGC18}  employs a RNN to sequentially encode the top results using their feature vectors and learn a local context model to re-rank the top results. We apply DLCM to re-rank the ranking list of base models.

$\bullet$ \textbf{PRM}~\cite{DBLP:conf/recsys/PeiZZSLSWJGOP19} employs a Transformer structure to model the global relationships between any pair of items in the whole list. We apply \textsl{PRM-BASE} to re-rank base model's results.


\subsubsection{Results}
The overall results are shown in Table~\ref{tab:baselines}, where we can make the following observations:

\begin{itemize}[leftmargin=*]
	\item[$\bullet$] In general, Ada-Ranker performs best on three datasets and with two base models (with an exception for GAUC metric in Taobao dataset), which demonstrates the effectiveness of our method. 
	\item[$\bullet$] DNS hurts the performances of base models, which indicates that our assumption -- DNS can encode and generalize to various types of data distribution -- does not hold. 
	\item[$\bullet$] Methods based on GSF explicitly model distribution of candidates through concatenation/average/Transformer, and the results show that these ways can boost GRU4Rec and SASRec. As for GSF$_{concat}$, although it achieves surprising performances on Taobao dataset, it is unstable on other two datasets. GSF$_{trm}$ almost outperforms GSF$_{avg}$ on all datasets, and this owes to the high-quality context-aware representations calculated by the multi-head self-attention. 
	\item[$\bullet$] PD and DecRS are causal recommendation models which revise the scores of candidates by considering data distribution bias. Both methods can improve the performance of GRU4Rec and SASRec. Note that PD performs best on Taobao dataset in terms of GAUC metric, but it is much worse than Ada-Ranker with NDCG on three datasets. To some extent, this result shows that PD is not robust to adapt model to a given list of candidates only by decoupling the user preference and data distribution bias. 
	\item[$\bullet$] DLCM and PRM are two re-ranking methods, which decouple the process of training base model and the post-processing of the ranking lists. However, the decoupled design cann't deeply fuse the base model with 
the candidate distribution information, so Ada-Ranker can outperform both DLCM and PRM.
	
\end{itemize}

\begin{table}[t!]
	\newcommand{\tabincell}[2]{\begin{tabular}{@{}#1@{}}#2\end{tabular}}
	\small
	\caption{Results of ablation Study.} 
	\label{tab:ablation_study}
	\vspace{-0.15in}
	\setlength{\tabcolsep}{0.8mm}{
		\begin{tabular}{cc|cc|cc|cc} 
			\toprule
			\multicolumn{2}{c|}{\multirow{2} * {Methods}} & \multicolumn{2}{c}{ML10M} & \multicolumn{2}{|c}{Taobao} & \multicolumn{2}{|c}{Xbox} \\
			&  & GAUC & NDCG & GAUC & NDCG & GAUC & NDCG \\
			\midrule
			\multicolumn{2}{c|}{Base (GRU4Rec)} & 0.9187 & 0.7781  & 0.8734 & 0.7208 & 0.9388 & 0.8302\\
			\hline
			(1) & avg & 0.9238 & 0.7853  & 0.8792 & 0.7349 & 0.9423 & 0.8384\\ 
			\hline
			\multirow{3} * {(2)}
			& w/o FiLM & 0.9261	& 0.7899 & 0.8861 & 0.7463 & 0.9444&	0.8416\\ 
			& add\_bias  & 0.9273 & 0.7921  & 0.8876 & 0.7472 & 0.9437 & 0.8422 \\
			& diff $\gamma, \beta$ & 0.9260	& 0.7889 & 0.8881 & 0.7481 & 0.9429 & 0.8392   \\
			
			\hline
			\multirow{7} * {(3)}
			& w/o mem\_net  & 0.9243 & 0.7863 & 0.8855 & 0.7459 & 0.9407 & 0.8357 \\ 
			& w/o global\_para & 0.9209	& 0.7819 & 0.8879 & 0.7448 & 0.9437 & 0.8411  \\
			& free\_para & 0.9269 & 0.7907 & 0.8851 & 0.7445 & 0.9442 & 0.8408  \\
			& add\_bias (1 layer)  & 0.9257 & 0.7887 & 0.8843 & 0.7409 & 0.9438 & 0.8409 \\
			& add\_bias (2 layers) & 0.9263 & 0.7894 & 0.8849 & 0.7415 & 0.9433 & 0.8396 \\
			& \#slots $L$=5 & 0.9260 & 0.7896 & 0.8881 & 0.7483 & 0.9444 & 0.8417 \\
			& \#slots $L$=15 & 0.9268 & 0.7909 & 0.8880 & 0.7480 & 0.9447 & 0.8423 \\
			\hline
			\multicolumn{2}{c|}{Ada-Ranker} & \textbf{0.9279} & \textbf{0.7932} & \textbf{0.8888} & \textbf{0.7490} & \textbf{0.9451} & \textbf{0.8438} \\
			\bottomrule
		\end{tabular}
	}
\end{table}

\begin{table}[t!]
	\newcommand{\tabincell}[2]{\begin{tabular}{@{}#1@{}}#2\end{tabular}}
	\caption{Comparisons of different training strategies.}
	\small
	\label{tab:training}
	\vspace{-0.15in}
	\setlength{\tabcolsep}{0.4mm}{
		\begin{tabular}{c|c|cc|cc|cc} 
			\toprule
			\multicolumn{1}{c|}{\multirow{3} * {Methods}} & \multicolumn{1}{c|}{\multirow{3} * {\tabincell{c}{Training\\ Strategies}}} &\multicolumn{6}{|c}{GRU4Rec}\\
			\cline{3-8}
			& & \multicolumn{2}{|c|}{ML10M} & \multicolumn{2}{|c|}{Taobao} & \multicolumn{2}{|c}{Xbox}\\
			
			&  & GAUC & NDCG & GAUC & NDCG & GAUC & NDCG \\
			\midrule
			
			Base & Train $\Theta$ & 0.9187 & 0.7781 & 0.8734 & 0.7208 & 0.9388 & 0.8302 \\
			\hline
			\multirow{3} * {Ada-Ranker} 
			& $\emptyset \Rightarrow \Theta + \Phi$ & 0.9279 & 0.7905 & 0.8805 & 0.7314 & 0.9434 & 0.8399 \\
			& $\Theta \Rightarrow \Theta + \Phi$  & \textbf{0.9288} & \textbf{0.7943} & \textbf{0.8906} & 0.7462 & \textbf{0.9488} & \textbf{0.8511} \\
			& $\Theta \Rightarrow  \Phi$  & 0.9279 & 0.7932 & 0.8888 & \textbf{0.7490} & 0.9451 & 0.8438 \\
			\bottomrule
		\end{tabular}
	}
\end{table}

\subsection{Ablation Study} 
\subsubsection{Effect of the Components}
\label{exp:ablation}
Key components in Ada-Ranker include (1) distribution extractor, (2) input modulation, and (3) parameter modulation. To verify each component’s impact, we disable one component or replace it with some variant each time while keeping the other settings unchanged, then test how the performance will be affected. The result is reported in Table ~\ref{tab:ablation_study}. 
\begin{itemize}[leftmargin=*]
	\item For (1), we replace the neural process (Section~\ref{sec:bias_ext}) by averaging all candidates' embeddings (denoted as \underline{avg}). The decline of performance indicates it is hard to fully capture the distribution information of the candidate list by simply averaging. Moreover, we provide visual analysis in Section \ref{sec:case_study}.
	\item For (2), we try different variants: remove the input modulation (\underline{w/o FiLM}); or replace FiLM with element-wise addition parameter (\underline{add\_bias}), which means that replacing Eq.(\ref{eq:im_2}) and Eq.(\ref{eq:im_1}) with $\tilde{\mathbf{q}}_t = \mathbf{q}_t + \tilde{\mathbf{b}}$ and $\tilde{\mathbf{b}} = f_b'(\mathbf{z})$; or let FiLM generate different $\gamma, \beta$ variables for each item in the item sequence (\underline{diff $\gamma,\beta$}), by taking item representation as additive input in Eq.(\ref{eq:im_1}). Removing FiLM block leads to a performance drop, which verifies the necessity of input modulation.   \textsl{diff $\gamma,\beta$} does not performs well, which shows that learning different affine transformations on each item in the sequence is difficult. \textsl{add\_bias}'s performance on Xbox dataset is poor. In comparison, FiLM's performance is robust and consistently good.
	\item For (3), we consider these variants: removing parameter modulation (\underline{w/o mem\_net});  removing global parameters of predictive layer (\underline{w/o global\_para}), which means replacing Eq.(\ref{eq:pm_modulate}) with $\widetilde{\mathbf{W}}_k =   \widehat{\mathbf{W}}_k, \ \ \widetilde{\mathbf{b}}_k =  \widehat{\mathbf{b}}_k$; generating adaptation parameters by Eq.(\ref{eq:pm_mlp}) (\underline{free\_para});  adding distribution bias vector after a linear projection to the first layer of prediction layer (\underline{add\_bias (1 layer)}) or to both layers (\underline{add\_bias (2 layers)});  changing the number of slots in memory network (\underline{\#slots $L=5$}, and \underline{\#slots $L=15$}). No matter removing memory network or replacing it with other configurations, the performances drop much compared with Ada-Ranker. 
\end{itemize}

\subsubsection{Effect of Different Training Strategies}\label{sec:traing_method}
We compare different training strategies (see Section~\ref{sec:parameter}) on \textbf{global parameters $\Theta$} and \textbf{adaptation parameters $\Phi$} in Ada-Ranker. Table~\ref{tab:training} shows that all methods can boost base models significantly. Specially, jointly training $\Theta$ and $\Phi$ from scratch performs worse than the other two methods in most cases, which shows that the two-stage training procedure is of great importance. Besides, fine-tuning both $\Theta$ and $\Phi$ outperforms fine-tuning $\Phi$ only (except a result in NDCG metric on Taobao dataset). This is in expectation.  $\Theta$ has far more parameters than $\Phi$. Making $\Theta$ fixed is the finetuning stage is a strict constraint, which limits the expressive ability to some extent.  However, as discussion in Section \ref{sec:parameter}, only fine-tuning $\Phi$ makes Ada-Ranker more flexible for real-world applications. 

\subsection{Qualitative Study}\label{sec:case_study}
\begin{figure}[t!]
	\centering
	\includegraphics[width=9cm]{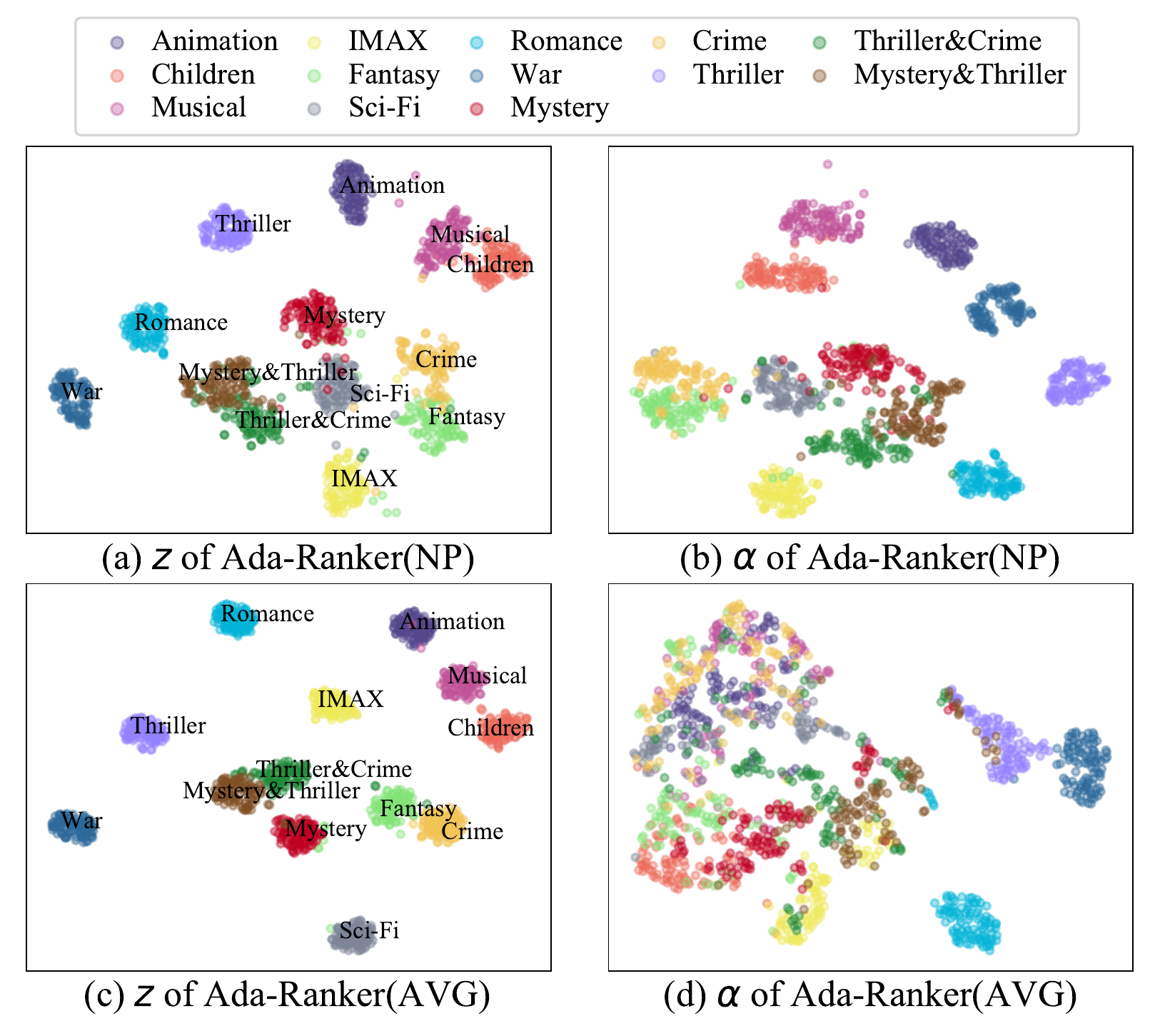}
	\vspace{-0.15in}
	\caption{t-SNE plots for distribution representation $\mathbf{z}$  (a and c) and coefficients $\boldsymbol{\alpha}$ of memory network (b and d), based on two different types of distribution extractor: NP and AVG. The dataset is ML10M.}
	\label{fig:case} 
\end{figure}

In this section, we explore whether our Ada-Ranker can distinguish different distributions of candidates. We choose 11 single categories (such as \textsl{Animation}, \textsl{Children}) and 2 mixed categories (including \textsl{Mystery\&Thriller} and \textsl{Thriller\&Crime}) of ML10M dataset. For each category, we generate 100 groups of candidate lists in which the items are all from the current category. For mixed category, we sample items from two categories with the proportion of 50\% and 50\%, respectively. 
Then, we apply the trained Ada-Ranker to infer on all candidate lists.
Figure~\ref{fig:case} shows t-SNE plots of the distribution vector $\mathbf{z}$ in Eq.~(\ref{eq:repara}) and the coefficients $\boldsymbol{\alpha}$ of memory network in Eq.~(\ref{eq:alpha_softmax}). We compare two types of Ada-Ranker: generating $\mathbf{z}$ by neural process (denoted as \underline{NP}) and by averaging all candidates embeddings (denoted as \underline{AVG}). Each color represents a category. We have the following observations:

\begin{itemize}[leftmargin=*]
	\item In Figure~\ref{fig:case}(a) and Figure~\ref{fig:case}(c),  distributions $\mathbf{z}$ generated by both NP or AVG can be clearly separated. However, we observe that $\mathbf{z}$ generated from NP has more flexibility.  For example, in terms of mixed distribution like \textsl{Thriller\&Crime}, the AVG method simply places $\mathbf{z}$-\textsl{Thriller\&Crime} in the middle of $\mathbf{z}$-\textsl{Thriller} and $\mathbf{z}$-\textsl{Crime}, while the NP method can locate $\mathbf{z}$-\textsl{Thriller\&Crime} to a better place to avoid overlapping with irrelevant categories. Similar conclusions can be drawn for $\mathbf{z}$-\textsl{Mystery\&Thriller}.  
	\item For parameter modulation, the AVG method fails to distinguish $\boldsymbol{\alpha}$ in memory network effectively (see Figure~\ref{fig:case} (d)). It shows   that our neural process encoder can model diverse distributions of different candidate lists well and improve the process of subsequent parameter modulation.
\end{itemize}

\subsection{Adaptation to New Distributions}\label{sec:new_distribution}
In Section \ref{sec:setting01} we introduce the distribution-mixer sampling as our default experiment setting, which ensures that both training set and test set contain diverse data distributions across different ranking requests. However, the prior distribution (i.e., Step 1 and Step 2 in the distribution-mixer sampling, which determine the distribution of item categories and item popularity) for both training set and test set are the same, which is a strong assumption and causes some gap between research and practice. In this section, we would like to answer this question: is Ada-Ranker generalized sufficiently to various scenarios,  even when the test set's prior distribution differs from the training set's (which means that we are handling new emerging requests)? 

\begin{figure}[t!]
	\centering
	\begin{subfigure}[b]{0.49\linewidth}
		\centering
		\includegraphics[width=\textwidth]{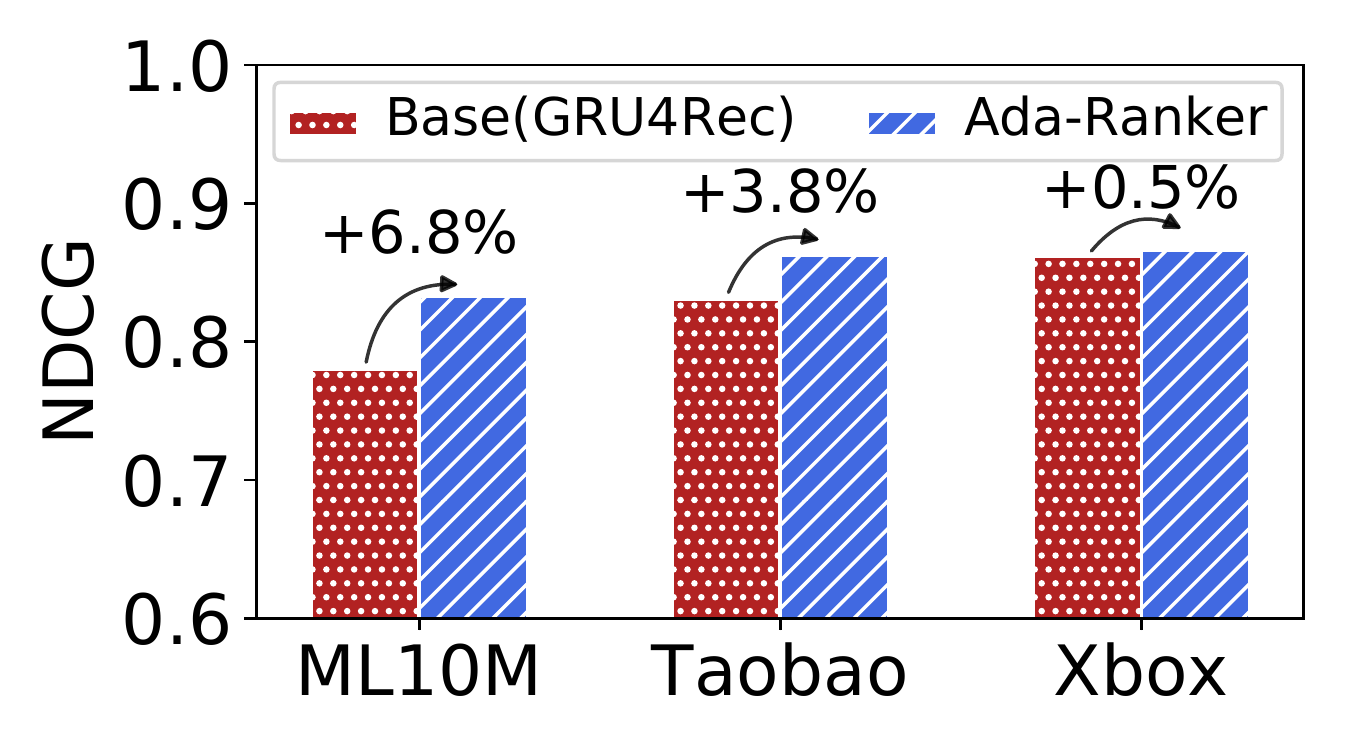}
		\vspace{-0.25in}
		\caption{GRU4Rec, NDCG, Same$_{Dis}$}
		\label{GRU4Rec_NDCG_1}
	\end{subfigure}
	\begin{subfigure}[b]{0.49\linewidth}
		\centering
		\includegraphics[width=\textwidth]{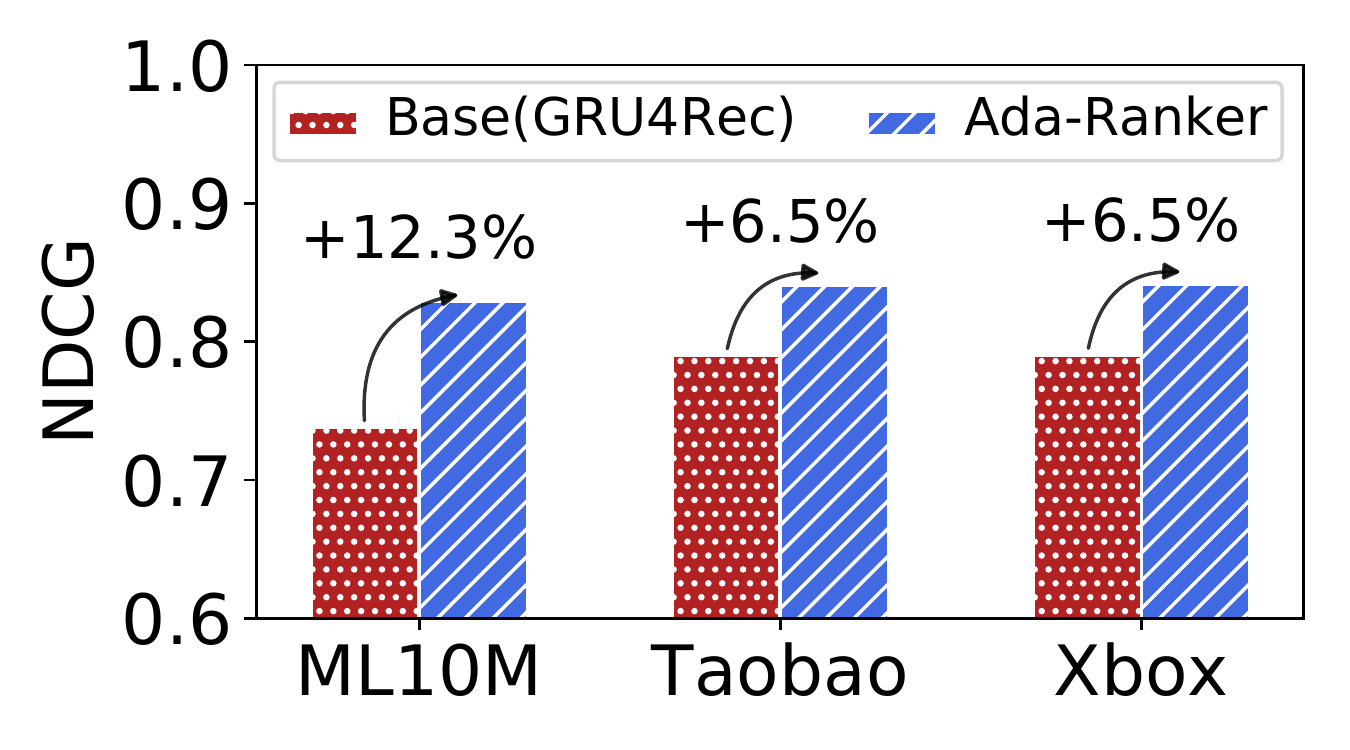}
		\vspace{-0.25in}
		\caption{GRU4Rec, NDCG, New$_{Dis}$}
		\label{GRU4Rec_NDCG_0}
	\end{subfigure}
	
	\begin{subfigure}[b]{0.49\linewidth}
		\centering
		\includegraphics[width=\textwidth]{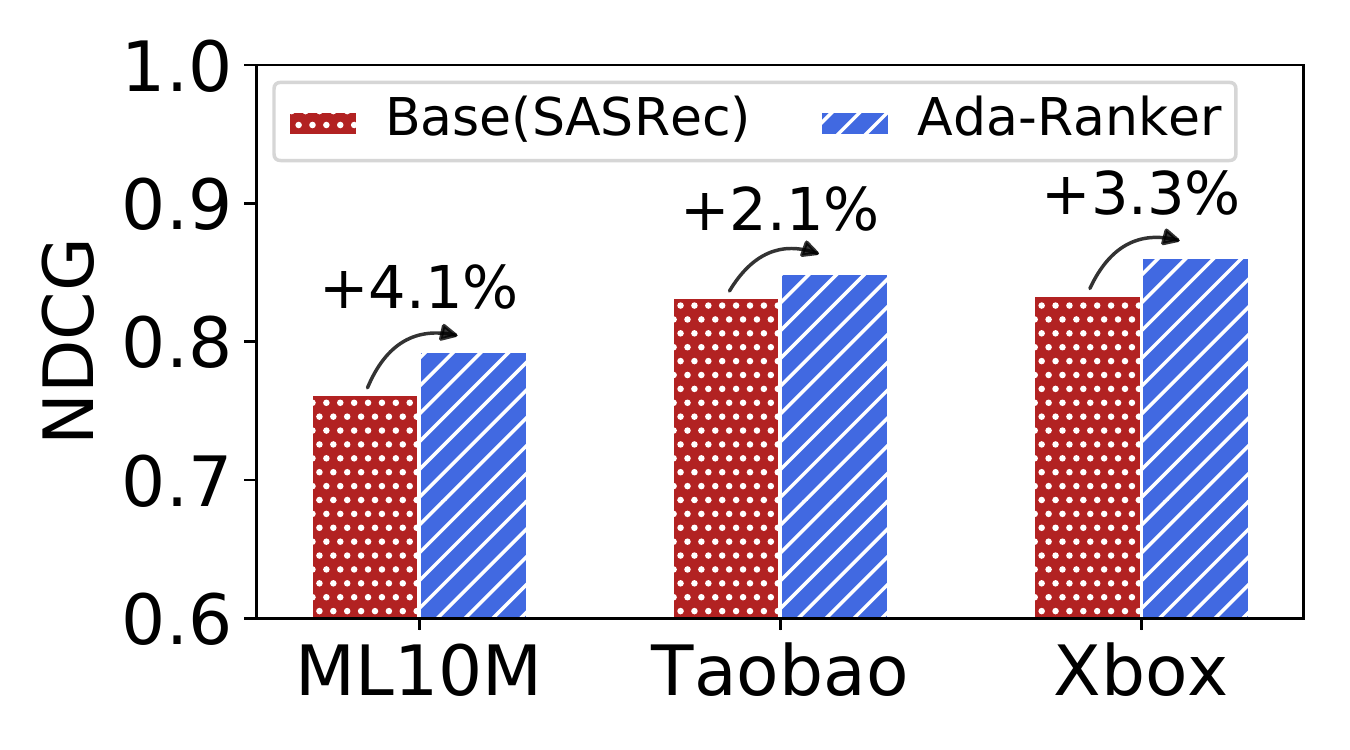}
		\vspace{-0.25in}
		\caption{SASRec, NDCG, Same$_{Dis}$}
		\label{SASRec_NDCG_1}
	\end{subfigure}
	\begin{subfigure}[b]{0.49\linewidth}
		\centering
		\includegraphics[width=\textwidth]{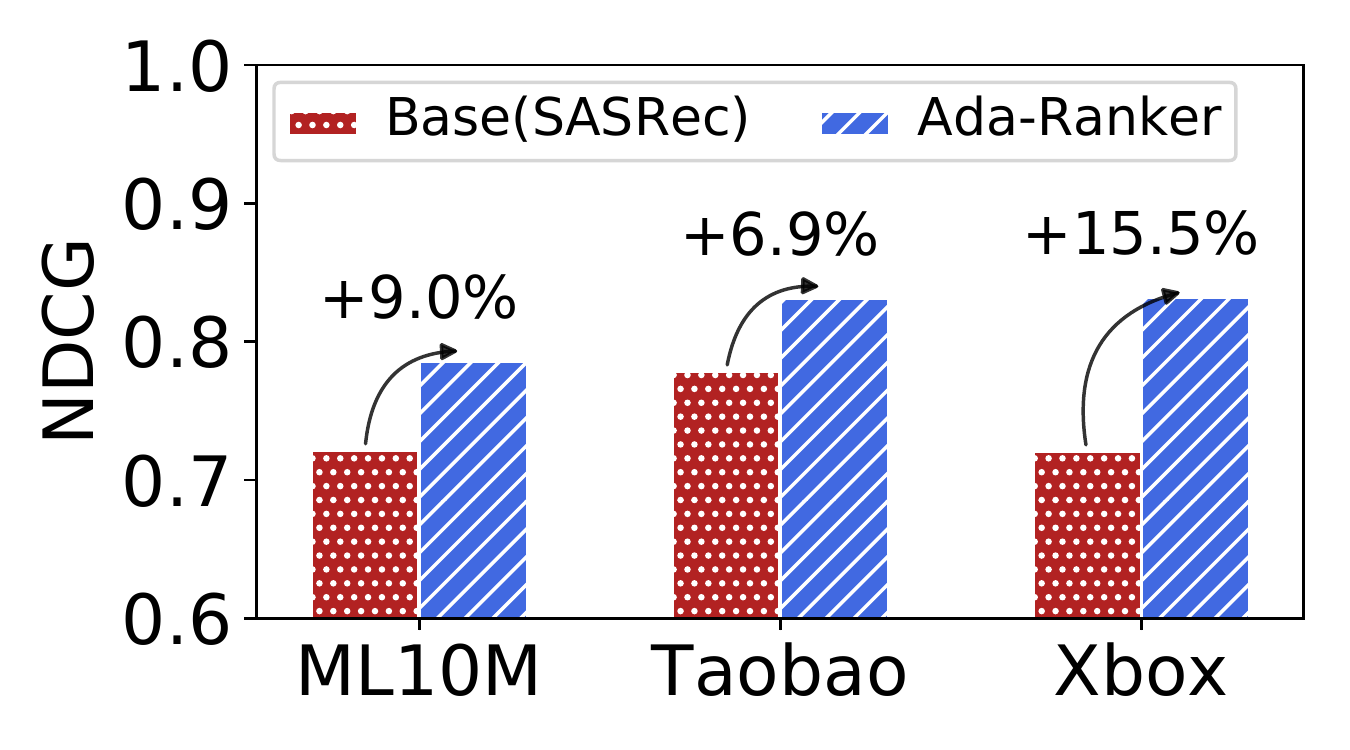}
		\vspace{-0.25in}
		\caption{SASRec, NDCG, New$_{Dis}$}
		\label{SASRec_NDCG_0}
	\end{subfigure}
	\label{MF}
	
	\caption{NDCG scores of Ada-Ranker with two base models \{GRU4Rec, SASRec\} on three datasets \{ML10M, Taobao, Xbox\} under different distribution settings. Same$_{Dis}$ means test set is under the same distribution as training set. New$_{Dis}$ means test set is generated from a new data distribution.}
	\label{fig:oof}
\end{figure}

\subsubsection{Settings}
We revisit the classical recall-and-rank setting. The candidates that a ranker model need to handle are proposed by some recall models. However, for an industrial recommender system, there are several reasons to cause the inference distribution not identical to the training distribution, such as (1) the intermediate candidate set from recall models are not displayed to users, so we are not able to reconstruct the recall distribution from impression logs; (2) the recall models themselves are evolving from time to time. Therefore, we design another kind of setting, where the major difference lies in negative instances selection, with the same positive instances in Section \ref{sec:setting01}.

We choose three widely-used recall methods (Popular, MF and Item2Item) to generate negative instances. 
Specifically, when constructing the training set, for each positive instance we generate 19 negative samples from the three recall models with a budget allocation of 20\%, 50\% and 30\%, respectively. Then, we generate two different test sets in order to compare the performances of base models and Ada-Ranker under the same data distribution or a new data distribution: (1) Same$_{Dis}$: sampling negatives with the same distribution as the training set, i.e., 20\%, 50\% and 30\% from the three recall methods. (2) New$_{Dis}$: sampling negatives with new distribution, i.e., sample negative items from Popular, MF, Item2Item with the proportion of 40\%, 10\% and 50\%, respectively. The second setting increases the difficulty of inference which is closer to the actual scenario. For more details please refer to the Appendix~\ref{app:new_dis_setting}.

\subsubsection{Results}
From Figure~\ref{fig:oof} and Figure~\ref{fig:oof_gauc}, we can observe that no matter the data distribution between training set and test set is the same or different, Ada-Ranker can improve the performance of two base models (GRU4Rec and SASRec) effectively and consistently, on three datasets and two metrics (results on GAUC metric is reported in Figure~\ref{fig:oof_gauc}). Besides, when the distribution of test set is new, the performances of base models drop a lot compared with the setting under the same data distribution (e.g., GRU4Rec performs 0.78 on ML10M dataset over NDCG metric under Same$_{Dis}$, while it only performs 0.73 on the new test set). Such a performance decrease is mainly caused by the inconsistent data distribution. Therefore, the relative improvement of the Ada-Ranker over the base model is more significant in the New$_{Dis}$ setting than in the Same$_{Dis}$ setting. For example, in the Xbox dataset, with SASRec as the base model, Ada-Ranker's NDCG gain in Same$_{Dis}$ is 3.3\%, while in New$_{Dis}$, it is as high as 15.5\%. This observation indicates that Ada-Ranker is capable of modeling the distribution of candidate samples and adjust parameters of the network accordingly, which alleviates the issue of inconsistent data distribution.

\section{related works} 

\paratitle{Neural Recommender Systems}. 
A web-scale recommender system usually comprises of recall modules and ranking modules to response users' requests in (near) real-time \cite{10.1145/2959100.2959190,ijcai2018-529}. Recall modules propose a small set (such as thousands) of potentially relevant items as candidates with lightweight models. Ranking modules will score the given item candidates by adopting  more advanced models to ensure the final recommendations are accurate. 

One important recommendation scenario is the sequential recommendation, where DNNs have been successfully applied, such as GRU4Rec~\cite{DBLP:journals/corr/HidasiKBT15}, NextItNet~\cite{10.1145/3289600.3290975}, and SASRec~\cite{DBLP:conf/icdm/KangM18}. However, all these approaches follow the same traditional train-freeze-test paradigm, where a global optimal model is learned from the training dataset and then is frozen  for testing. The testing environment in recommender systems is highly dynamic with diverse users' requests, thus, it needs an approach that can perceive the current task's peculiarity and conduct suitable parameter adaptation.

\paratitle{Domain Generalization}.
Domain generalization has attracted increasing interests in many application fields (e.g., NLP and CV). It aims at learning a model from several different but related domains, then generalize to unseen testing domain \cite{wang2021generalizing}. Related techniques can be roughly divided into four categories.

\emph{Multi-task Learning}: 
Industrial recommender systems usually serve diverse application situations. For example, the \textsl{banner} and \textsl{Guess What You Like} channels on Taobao app homepage~\cite{sheng2021model}. Multi-task learning (MTL) optimizes a model by jointly training several related tasks simultaneously. Typical models include  parameter sharing of hidden layers~\cite{DBLP:books/sp/98/Caruana98}, Multi-gate Mixture-of-Experts (MMoE)-based methods ~\cite{DBLP:conf/kdd/MaZYCHC18,10.1145/3383313.3412236}, and Star Topology Adaptive Recommender (STAR)~\cite{sheng2021model}. 
Note that all these MLT methods can only work when the tasks/domains are fixed and well-defined, they can hardly generalize to new domains or ad-hoc tasks with dynamic distributions.

\emph{Model Patch}: 
Finetuning a pre-trained model with target task-specific data is the easiest way to transfer knowledge from source domains to a target domain. In this line of research, this work~\cite{DBLP:conf/icml/HoulsbyGJMLGAG19} proposes a parameter-efficient tuning paradigm called \textsl{model patch}, which adapts a large-scale pre-trained model to a specific downstream task by finetuning only a small number of new parameters.
PeterRec~\cite{10.1145/3397271.3401156} aims to adapt a user representation model for multiple downstream tasks. 
All these models require that the tasks/domains are fixed and known beforehand, they cannot directly apply to our scenario, but the idea motivates the design of our Ada-Ranker.

\emph{Meta-learning}:  
Meta-learning is a learning-to-learn paradigm, aiming to learn experience from a distribution of related tasks so as to improve its future learning performance~\cite{hospedales2020meta}. In particular, MAML~\cite{finn2017model} based methods is frequently applied recently in cold-start recommendations for fast adaptation to new users~\cite{9338389,lee2019melu,yu2020personalized,du2019sequential}. However, it requires a support set with supervised labels in the adaptation stage, which is not feasible in our setting, where all the items in candidate set $\mathcal{C}$ are unlabeled.

\emph{Feature Modulation}:  
Another popular approach to adaptively altering a neural network's behavior with dynamic conditions is FiLM~\cite{DBLP:conf/aaai/PerezSVDC18}, which carries out a feature-wise linear modulation on a neural network's intermediate features conditioned on an arbitrary given query, so that the neural network computation over an input can be influenced by a dynamic condition. TaNP~\cite{10.1145/3442381.3449908} introduces an improved version of FiLM, named Gating-FiLM, to enhance the ability of filtering the information which has the opposite effects during adaptation.  AdaSpeech~\cite{DBLP:conf/iclr/Chen0LLQZL21} adapts few adaptation data to personalized voices, which modulates hidden state vectors in network by categorizing the acoustic conditions in different levels and conditional layer normalization. In this paper, FiLM serves as one of the fundamental components in Ada-Ranker.

\paratitle{List-wise Ranking and Re-Ranking}.
List-wise learning-to-rank methods, which assign scores to items in a list by maximizing the utility of the entire list, are also related to our work. Most of them train models with list-wise loss functions, however, in the inference stage, they still use a point-wise scoring function (aka univariate scoring function) which is unaware of the context of candidate list~\cite{burges2006learning,cao2007learning,burges2010ranknet}. GSFs~\cite{DBLP:conf/ictir/AiWBGBN19} proposes multivariate scoring functions, in which the score of an item is determined jointly by multiple items in the candidate list. However, GSFs is hard to converge to a satisfying result because that by jointly modeling all items in the list, a large amount  of new parameters are involved. Another line of related research is re-ranking~\cite{AiBGC18,ijcai2018-518,DBLP:conf/recsys/PeiZZSLSWJGOP19}, which refines a ranker's suggested top-$k$ items with another model by taking the top top-$k$ items as ranking context. Typically, they are post-processing methods which are decoupled with ranker models. Since they take the whole candidate items as inputs, we can expect that they naturally have the ability to capture data distributions in candidate set. So we use them as baselines in Section~\ref{sec:baselines}.


\section{Conclusion}
In this paper, we propose a novel ranking paradigm, Ada-Ranker, which can perceive the data distribution of a specific candidate list and learn to adjust the ranker model before making predictions. In contrast to traditional rankers' \textsl{train-freeze-test} paradigm, Ada-Ranker uses a new \textsl{train-adapt-test} paradigm. Ada-Ranker is lightweight, model-agnostic, and flexible for plug-and-play usage.  
We have conducted extensive experiments on three real-world datasets. Results show that Ada-Ranker can effectively enhance various types of base sequential recommender models, as well as outperform a comprehensive set of competitive baselines.

\begin{acks}
This work was partially supported by National Natural Science Foundation of China under Grant No. 61872369,
Beijing Natural Science Foundation under Grant No. 4222027,  and Beijing Outstanding Young Scientist Program under Grant No. BJJWZYJH012019100020098. This work is also partially supported by Beijing Academy of Artificial Intelligence (BAAI). Xin Zhao is the corresponding author.
\end{acks}

\bibliographystyle{ACM-Reference-Format}
\bibliography{sample-base}

\appendix
\section{appendix}

\renewcommand\thefigure{\Alph{section}\arabic{figure}}
\renewcommand\thetable{\Alph{section}\arabic{table}}
\subsection{Settings of sampling negative instances from recall models}\label{appendix}
\setcounter{figure}{0}
\setcounter{table}{0}

In section~\ref{sec:new_distribution}, we briefly introduce our method of constructing test sets with different distributions. In this section, we will introduce more details to help readers understand how we sample negative instances from recall models.

\subsubsection{Three Recall Models}
We choose Pop, MF, Item2Item as basic recall models to generate negative instances.
\begin{itemize}[leftmargin=*]
	\item[$\bullet$] \textbf{Pop}. We recall 1000 candidates from all items according to their popularity.
	\item[$\bullet$] \textbf{MF}. We first train a simple collaborative filtering model (e.g., MF) and obtain the user embedding table and the item embedding table. For each user, we will product its embedding with all items' representations, and recall candidates ranked between 1000 and 2000.
	\item[$\bullet$] \textbf{Item2Item}. We take the historical behaviors of users as training data, and train item embeddings with the word2vec algorithm~\footnote{https://radimrehurek.com/gensim/models/word2vec.html}. For each item, we will product its embedding with all items' representations, and recall candidates ranked between 500 and 1500.
	
\end{itemize}

\subsubsection{Sampling According to Different Distributions}
\label{app:new_dis_setting}
We assume the distribution of sampling is $\mathbf{d}=[d_1, d_2, d_3]$, where $\sum_{i=1}^3 d_i = 1.0$. Each element in $\mathbf{d}$ represents the probability of sampling from the three recall models (Pop, MF and Item2Item, respectively).

For each positive instance $(u, v)$, we firstly use the function \textit{numpy.random.multinomial} to obtain an instantiated number of items to be sampled by each recall model: $\mathbf{x}=[x_1, x_2, x_3]$, with $\sum_{i=1}^{3}x_i=19$. Then, we carry out the following three steps to generate the whole 19 negatives. 
\begin{itemize} 
\item  Randomly sample $x_1$ items from the candidates in the range [0,  1000] sorted by the item's popularity score. 
\item  Randomly sample $x_2$ items from the candidates in range [1000, 2000]  sorted by the MF score (item $i$'s score is the dot product of embedding vectors of user $u$ and item $i$). 
\item  Randomly select one of the items that user has recently interacted with, say $t$. Randomly sample $x_3$ items from the candidates  in the range [500, 1500] sorted by the Item2Item score (item $i$'s score is the dot product of embedding vectors of item $t$ and item $i$, and here the embedding vector is not from MF, but from a word2vec algorithm). 
\end{itemize}

For training set, we set $\textbf{d}$ as $[0.2, 0.5, 0.3]$. For test set, we designed two settings: Same$_{Dis}$ ($\textbf{d}=[0.2, 0.5, 0.3]$) and New$_{Dis}$ ($\textbf{d}=[0.4, 0.1, 0.5]$). In section~\ref{sec:new_distribution}, we train the base models and Ada-Ranker on the training set, and infer on two different test sets.

\subsection{Additional experimental results for Section~\ref{sec:new_distribution}} 
\begin{figure}[h!]
	\centering
	\begin{subfigure}[b]{0.49\linewidth}
		\centering
		\includegraphics[width=\textwidth]{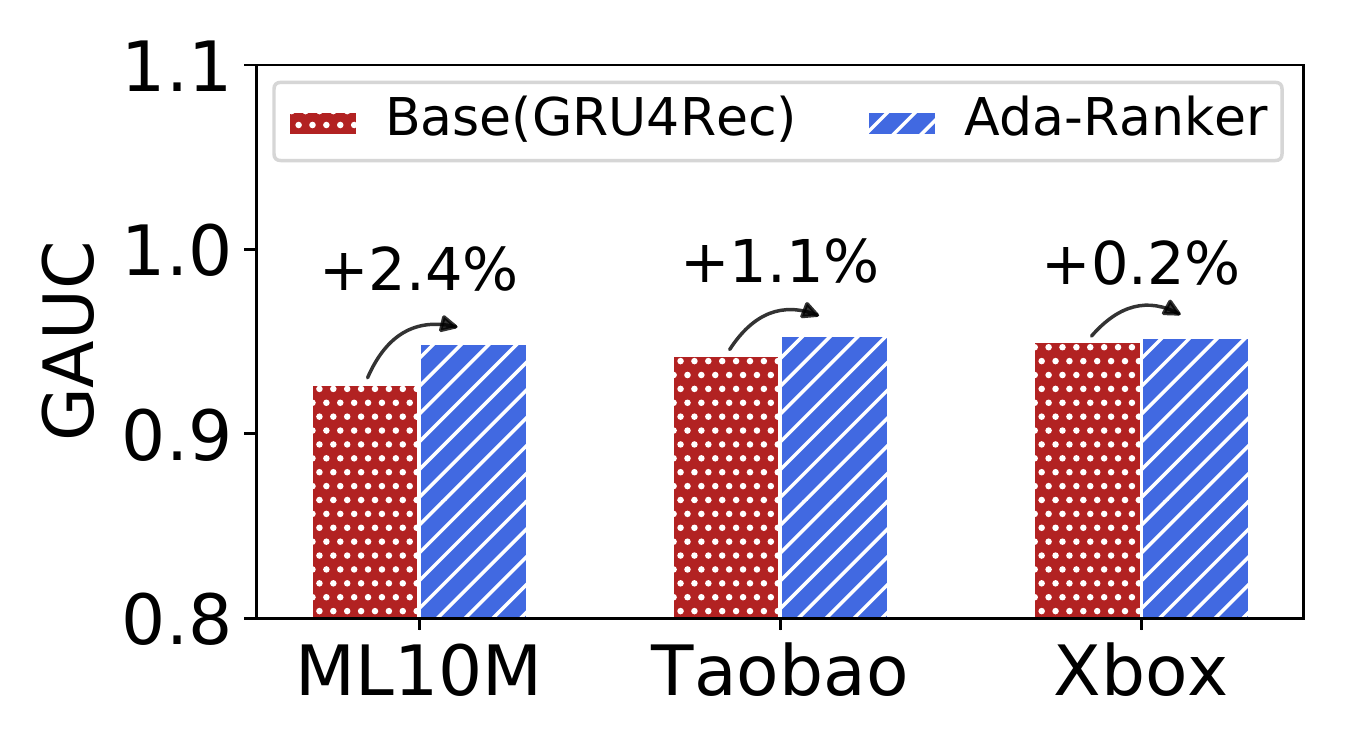}
		\vspace{-0.25in}
		\caption{GRU4Rec, GAUC, Same$_{Dis}$}
		\label{GRU4Rec_GAUC_1}
	\end{subfigure}
	\begin{subfigure}[b]{0.49\linewidth}
		\centering
		\includegraphics[width=\textwidth]{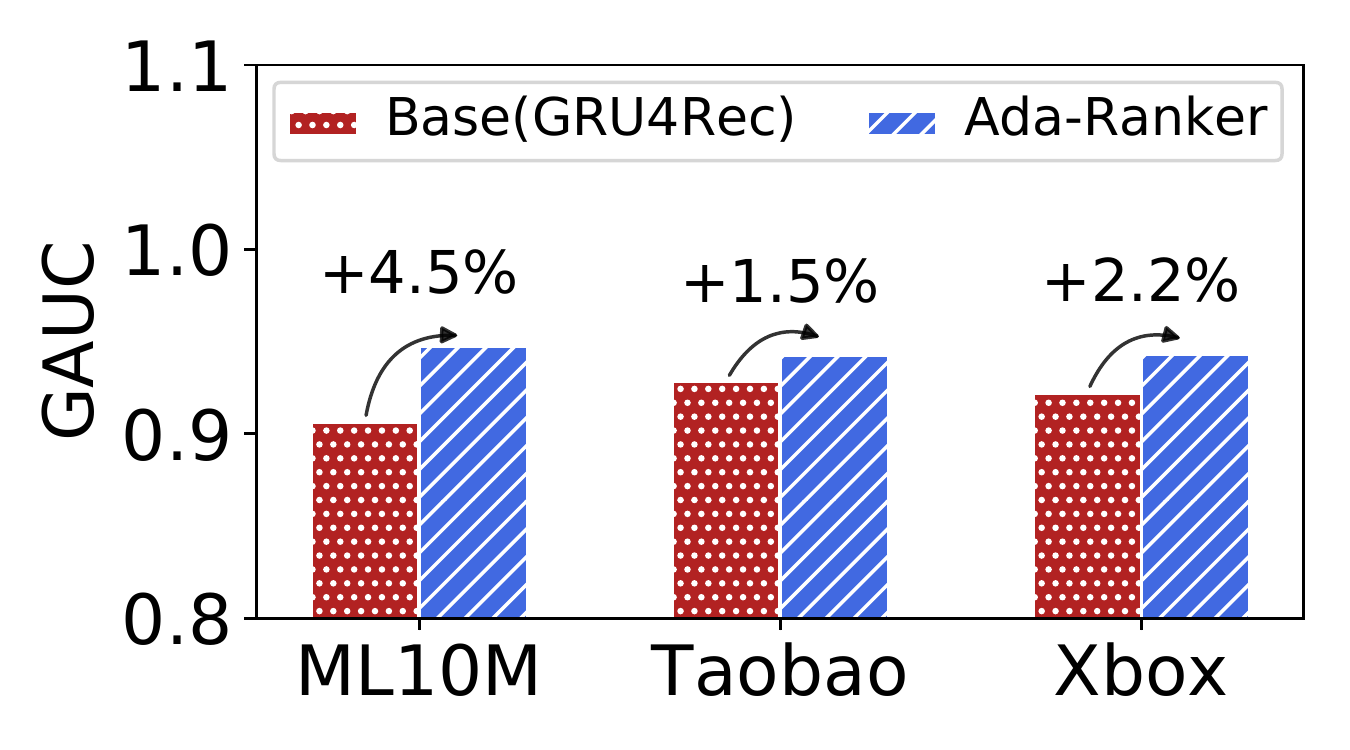}
		\vspace{-0.25in}
		\caption{GRU4Rec, GAUC, New$_{Dis}$}
		\label{GRU4Rec_GAUC_0}
	\end{subfigure}
	
	\begin{subfigure}[b]{0.49\linewidth}
		\centering
		\includegraphics[width=\textwidth]{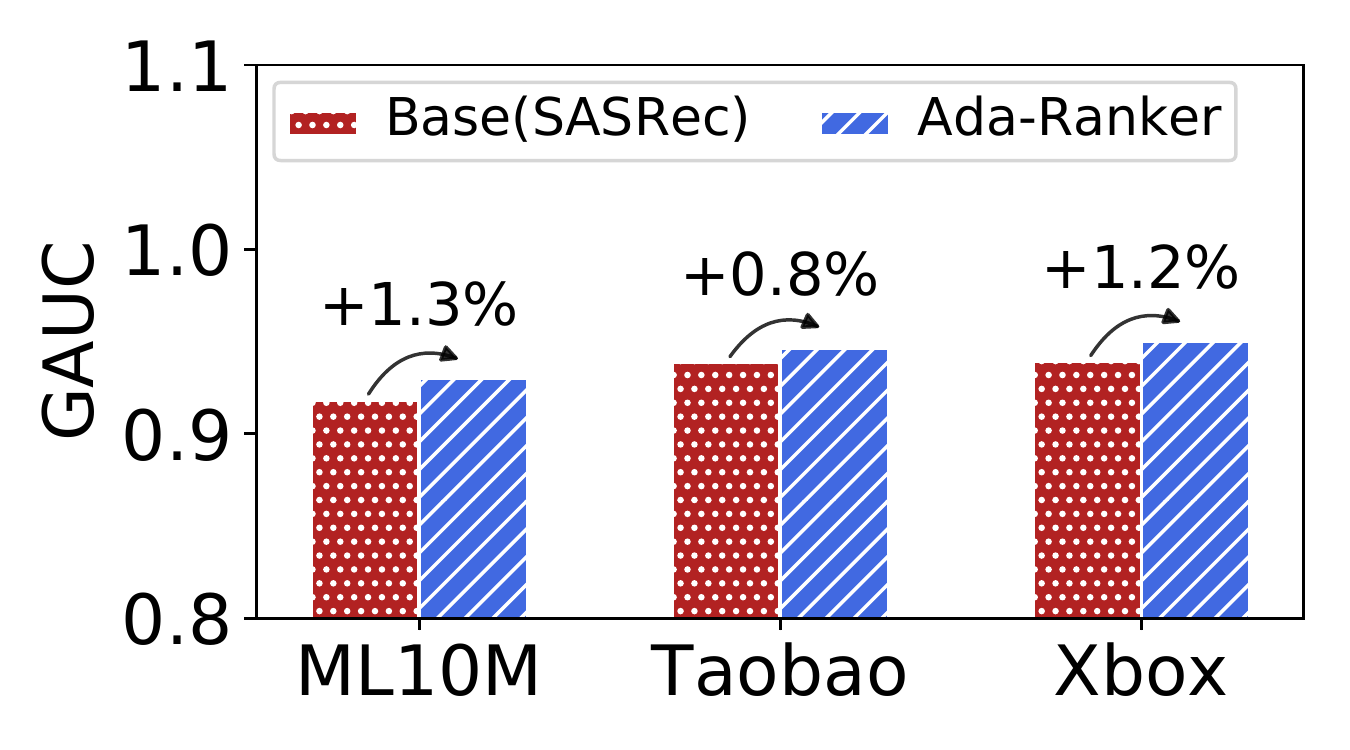}
		\vspace{-0.25in}
		\caption{SASRec, GAUC, Same$_{Dis}$}
		\label{SASRec_GAUC_1}
	\end{subfigure}
	\begin{subfigure}[b]{0.49\linewidth}
		\centering
		\includegraphics[width=\textwidth]{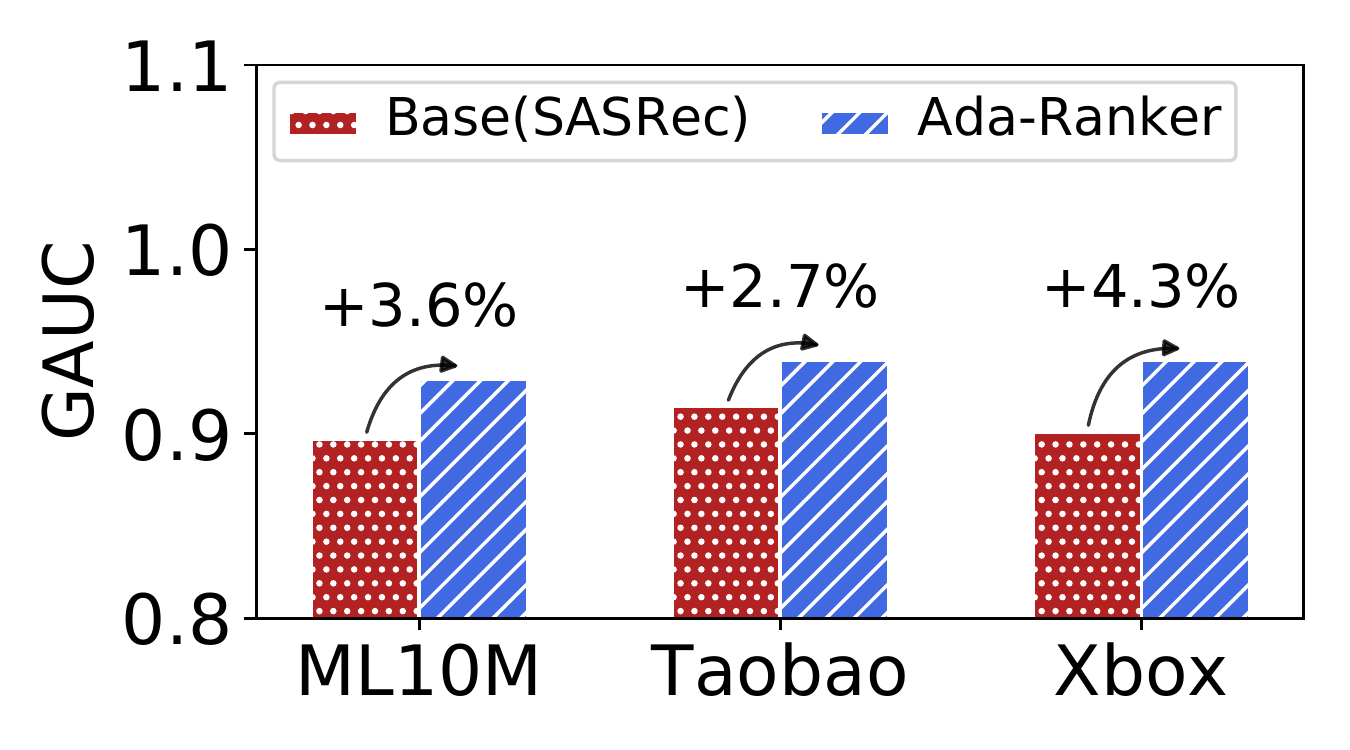}
		\vspace{-0.25in}
		\caption{SASRec, GAUC, New$_{Dis}$}
		\label{SASRec_GAUC_0}
	\end{subfigure}
	\label{GRU4Rec}

	\caption{GAUC scores of Ada-Ranker with two base models \{GRU4Rec, SASRec\} on three datasets \{ML10M, Taobao, Xbox\} under different distribution settings. Same$_{Dis}$ means test set is under the same distribution as training set. New$_{Dis}$ means test set is generated from a new data distribution.}
	\label{fig:oof_gauc}
\end{figure}

See Figure~\ref{fig:oof_gauc}.

\end{document}